%% file: xpc-hybrid.tex
\documentclass[10pt,emptycopyrightspace]{ewsn-proc}

\usepackage{graphicx}
\usepackage{balance}
\usepackage{comment}

\usepackage{cite}
\usepackage{verbatim}
\usepackage{amsmath,amssymb,amsfonts}
\usepackage{subcaption}
\usepackage{textcomp}
\usepackage{xcolor}
\usepackage{multirow}
\usepackage{booktabs}
\usepackage{adjustbox}

\usepackage{verbatim}

\usepackage{mathtools}

\usepackage{algorithm}
\usepackage[noend]{algpseudocode}

\newcommand{\vars}{\texttt}

\let\oldReturn\Return
\renewcommand{\Return}{\State\oldReturn}

\usepackage{array}
\newcolumntype{?}{!{\vrule width 1pt}}

\algnewcommand\algorithmicswitch{\textbf{switch}}
\algnewcommand\algorithmiccase{\textbf{case}}
\algnewcommand\algorithmicassert{\texttt{assert}}
\algnewcommand\Assert[1]{\State \algorithmicassert(#1)}%

\algdef{SE}[SWITCH]{Switch}{EndSwitch}[1]{\algorithmicswitch\ #1\ \algorithmicdo}{\algorithmicend\ \algorithmicswitch}%
\algdef{SE}[CASE]{Case}{EndCase}[1]{\algorithmiccase\ #1}{\algorithmicend\ \algorithmiccase}%
\algtext*{EndSwitch}%
\algtext*{EndCase}%

\numberofauthors{4}
\author{Alberto Spina, Michael Breza, Naranker Dulay, Julie McCann \\
	Department of Computing, Imperial College, London \\
	email: [alberto.spina15, mjb04, n.dulay, jamm]@imperial.ac.uk
}

\title{XPC: Fast and Reliable Synchronous Transmission Protocols for 2-Phase Commit and
3-Phase Commit}

\begin{document}

% make the title

\maketitle

\begin{abstract}
%Version that includes programming abstraction XPC
One of the major challenges for the engineering of wireless sensing systems is to
improve the software abstractions and frameworks that are available to
programmers while ensuring system reliability and efficiency. The distributed
systems community have developed a rich set of such abstractions for building
dependable distributed systems connected using wired networks, however after 20 years research many of these elude wireless sensor systems. In this paper we
present X Process Commit (XPC) an atomic commit protocol framework that
utilizes Synchronous Transmission (ST). We also introduce  
\textit{Hybrid}, a technique that allows us to exploit the advantages of  
the Glossy and Chaos Synchronous Transmission primitives to get lower latency and
higher reliability than either.
Using XPC and \textit{Hybrid} we demonstrate how to build protocols for the classical 2-phase and 3-phase commit abstractions and evaluate these demonstrating significantly improved performance and reliability than 
the use of Glossy or Chaos individually as dissemination primitives.
We address how we overcame the timing challenges of bringing Glossy and Chaos together to form \textit{Hybrid} and through extensive experimentation demonstrate that it is robust to
in-network radio interference caused by multiple sources. We are first to present testbed results that show that Hybrid can provide almost 100\% 
reliability in a network of nodes suffering from various levels of radio interference. 

% as12015 C0.01:
% > "in a network of 22 nodes"
% This is incorrect for the latest test data.
% When testing for multiple interference we start from a network of 24 nodes and gradually take nodes off the network and turn them into jammers (so network size from 24->16), where 16 is in the case of 8 jammers which we might prefer not to show given unstable data that I have not fully finished to re-run the tests for (see email). Surely we can provide quality data for up to 5 jammers.

%Non-XPC version
%One of the big challenges for the engineering of wireless sensing systems such
%as Wireless Sensor Networks is to improve the abstractions and frameworks that
%are available to system developers while ensuring system reliability. The
%distributed systems community have several such abstractions for dependable
%distributed systems implemented on wired networks. In this paper we present a
%hybrid synchronous transmission scheme called \textit{Hybrid} for the implementation of
%2-phase and 3-phase commit protocols. \textit{Hybrid} is robust to in-network radio
%interference such as that caused by other users and machines that emit noise in
%the same frequency. We present results that show that \textit{Hybrid} can provide near
%100\% reliability with latencies in the range of 1300 milliseconds in a network
%of 22 nodes suffering from high radio interference.  We use XPC and \textit{Hybrid} to
%build and evaluate 2-phase and 3-phase commit protocols that offer better
%performance and reliability than A\textsuperscript{2}/Synchrotron.  

\end{abstract}

\input{Sections/Introduction.tex}

\input{Sections/Background_Motivation.tex}
\input{Sections/Design-Implementation.tex}

\input{Sections/Evaluation.tex}
%\input{Sections/Case_Study.tex}
\input{Sections/Discussion_Limitations_FurtherWork.tex}
\input{Sections/Conclusion.tex}

%\Section{Introduction}
%
%\Section{Hardware/Software models of IOT sensor actuator nodes}
%
%\Section{WSN testing background}
%
%\Section{Conclusion}

%
% NOTE
%
% The following commands are all you need in the
\balance
\bibliographystyle{abbrv}
\bibliography{bib}  % sigproc.bib is the name of the Bibliography in this case
\end{document}

%% file: Sections/Introduction.tex
\section{Introduction}
%mike don't put my ??? into comment until they are resolved

Wireless Sensor Networks(WSN) are a key technology in environmental and
infrastructure monitoring and are giving rise to new solutions across many
sectors of industry, including manufacturing, electricity, gas and water supply,
construction and agriculture 
%\cite{baggio2005wireless,stajano2010smart,shi2018parkcrowd,qin2019low} 
%\cite{baggio2005wireless,stajano2010smart,shi2018parkcrowd} 
\cite{baggio2005wireless,stajano2010smart} 
%???PUT SOME OF OUR WORK HERE <-- added newest that I could find on your publication page???.  
One of the big challenges is to improve the software abstractions and frameworks 
that are available to programmers to create full-featured  
%???CREATE, TOO VAGUE, NOT SELLING WHY THEY WOULD NEED A NEW PRIMATIVE, WHAT'S THE BIG PICTURE <-- added references to system level services??? 
WSN that provide system level services like reconfiguration and update, while ensuring efficiency and reliability. This is
difficult due to the limited capabilities of sensor nodes as well as the difficulty
of overcoming the (often extreme) physical environment impediments that sensors will be
presented with potentially causing high rates of both communication and node failure.

The distributed systems community have developed many high-level abstractions
for building dependable distributed systems such as for: reliable broadcast,
consensus, group membership and view-synchronous communication. The WSN
community on the other hand have mostly assumed an asynchronous model and used
simpler and less dependable abstractions such as best-effort communication and
flooding due to previous difficulties with reliable synchronisation and communication.
%???DUE TO???.  
Yet programmers can benefit from higher-level abstractions, such as the
ability to reach agreement, if these abstractions are efficient and reliable. An
example application would be to support system reconfiguration like the choice
of a new sample rate for sensing applications, perform coordinated in-network processing,
%???Ref too OLD <-- added ref from above, parkcrowd??
%\cite{baggio2005wireless,stajano2010smart,shi2018parkcrowd,kolcun2016efficient}  
\cite{baggio2005wireless,stajano2010smart,kolcun2016efficient}  
or to make the decision to update to a new software image \cite{langiuupkit}. 
%???WE NEED TO THINK OF MORE <-- Added Dragon for in network processing???

Synchronous Transmission (ST) is an important approach where wireless nodes
can synchronise and communicate at the same time \cite{baloo}. Simply put, the physics of
synchronous transmission is that if two identical messages arrive at a
receiver within $0.5\mu$seconds then the messages will constructively
interfere, and be received correctly. If two messages are different, and arrive
within $160\mu$seconds and have at least a $3dB$ difference in signal strength, then the
stronger signal will be successfully received in spite of the presence of
another signal for $2.4GHz$ radio communication. This is called the capture 
effect and is the result of non-destructive interference. 
%???ARE THE NUMBERS HERE FOR A PARTICULAR PIECE OF KIT/OR PARTICULAR TO A FREQUENCY -- IF SO SAY THIS???

Synchronous Transmission allows us to implement abstractions based on the
synchronous model for distributed systems, where there is a known upper bound
%???SPEED <-- message processing time???
on message transmission and processing time.
Glossy and Chaos are two well-known ST examples. Glossy is a reliable,
one-to-all, ST communication primitive \cite{glossy} while Chaos is an
unreliable, all-to-all, ST communication primitive \cite{chaos}. Chaos can
terminate much faster than Glossy, but its performance can suffer from network
instability and it can fail to terminate in some cases. Therefore a primitive that harnesses the reliability of Glossy with the speed of Chaos can underpin the abstractions we require.
%A\textsuperscript{2}/Synchrotron \cite{al2017network} is a recent
%framework that uses Chaos to implement ST protocols for the classical 2-phase
%commit and 3-phase commit abstractions.

%It has been shown to provide reliable communication in the
%presence of network interference. As Glossy is one-to-all, it can only handle
%communication serially.

% as12015 C1.01:
% > "22 nodes with multiple sources"
% See comment as12015 C0.01

This paper makes a number of contributions. We present XPC (X Process Commit), a new programming framework for
the implementation of atomic commit protocols, and \textit{Hybrid}, a novel ST
approach that uniquely uses the  
Glossy one-to-all ST primitive and the Chaos all-to-all ST primitive to achieve
better reliability and speed than either on their own. To achieve this \textit{Hybrid} is required to make 
decisions on the use of the appropriate ST primitive with the best parameters at the time.
%???ELUDE TO THE CHALLENGE TO GET THESE TOGETHER HERE IN SUMMARY -- TO GET THE READER EXCITED 
%ABOUT WHAT WE DID, THEN THEY'LL READ IN MORE DETAIL IN THE COMING PARAGRAPHS???
We provide a detailed evaluation and comparison of Glossy, Chaos, and \textit{Hybrid}
when used for the two-phase commit \cite{twopc} and three-phase commit \cite{threepc} protocols. 
Our results show that in a network of 20 nodes with two 
sources of high radio interference \textit{Hybrid} can provide close to 100\%
reliability when Chaos can not, and latencies that are between 13\% - 50\%
faster than Glossy.

%% file: Sections/Background_Motivation.tex
\section{Background and Related Work}

Wireless sensing systems are difficult to build because of the high rates of
failure of both their communication networks and their sensor nodes
\cite{breza2018failures}. Protocols and software abstractions are needed to transform
unreliable nodes and links into dependable wireless sensing systems that
provide services like network wide updates and in-network processing to support
a diverse range of application domains from smart cities to precision
agriculture.    
%???SO....???

\subsection{Atomic Commit Protocols} Atomic Commit Protocols
\cite{lynch1996distributed} are important to all distributed systems that need
to maintain a consistent global state across the entire system. Examples for WSN
systems might include the uniform rate of sampling for all of the sensors, and the
use of the same code version on all of the sensor nodes. Protocols for
2-phase commit (blocking) and 3-phase commit (non-blocking) are the most well established
and used to ensure that the nodes in a distributed system agree to commit a transaction,

\subsection{Synchronous Transmission}
Synchronous Transmission (ST) communication primitives aim to provide energy and time efficient network-wide
broadcasts by synchronously transmitting packets from multiple wireless nodes.
They depend upon the radio effects of constructive interference
\cite{chang2018constructive}, the capture effect \cite{gezer2010capture}, or
both. Constructive interference occurs when two identical radio messages are
received within $0.5 \mu$seconds of each other and can be successfully decoded.
The Glossy ST communication primitive \cite{glossy} was one of the first to use
constructive interference, followed by many others
\cite{chang2018constructive}. The requirement that both messages are similar makes
constructive interference based schemes inherently one-to-many. 

%The capture effect occurs when two messages, similar or different, are received
%within $160  \mu$seconds, and there is a received signal strength difference of
%at least $3dB$. ???DELETE AS ITS REPEATED FROM BEFORE???? In this case, the 
%first message will be successfullydecoded.
The Chaos communication primitive \cite{chaos} was one of the first examples of the use
of the capture effect. The
relaxation of the message similarity requirement of constructive interference
makes communication primitives using the capture effect all-to-all. 
%???DOUBLE CHECK THE USE OF THE WORDS 'THE FIRST' THAT INDEED THEY WERE THE FIRST <-- Changed???

The existence of communication redundancy makes ST communication very reliable
in practice. A notable exploration of this property has been the EWSN(Embedded
Wireless Systems and Networks conference) Dependability Competition that has
been held to assess the reliability of communication primitive , 
and propose a methodology to assess this \cite{boano2017ewsn}. The
use of ST protocols is being explored for many high reliability applications
\cite{baddeley2019atomic}.
\subsection{A\textsuperscript{2}/Synchrotron} \label{sec:a2-synchrotron} 

Synchrotron \cite{al2017network} is a transmission kernel inspired by Chaos and
LWB \cite{ferrari2012low}. It operates in time slots which include the time
taken for the reception, processing and transmission of packets. In Chaos,
reception rates degrade quickly where there is network interference and link
unreliability. Synchrotron addresses this by using time-slotted channel hopping,
%(TSCH) \cite{tsch}, 
it spreads transmissions across multiple channels. Each
node chooses one channel to use for transmissions within a given time slot.
This increases the protocol's reliability.

A\textsuperscript{2} is a programming library built on top of Synchrotron that
provides several higher-level abstractions \cite{al2017network}:

\begin{itemize}
    \item \textbf{Disseminate, collect and aggregate}. Allows for many-to-one
    and one-to-many communication. 
    \item \textbf{Vote}. Allows
    nodes to vote on a coordinator node's proposal. Voting is best-effort.
    \item \textbf{Agreement}. 2PC and 3PC for network-wide agreement.
    \item \textbf{Group membership}. Persistent groups with join and leave capabilities.
\end{itemize}

Synchrotron suffers from the same scalability and reliability problems as
Chaos. Chaos uses a control message that uses a single bit per node in the
network to keep track of which nodes have received the latest data. This limits
the size of the network that Chaos can be used on. 
%Chaos is also best effort,
%and provides no guarantees that all nodes will hear the latest data, or that
%the flood will terminate. Synchrotron addresses channel
%unreliability problems by using multiple channels.

%???ALL THE ABOVE TECH SHOULD BE CRITIQUED NOT JUST DESCRIBED, OTHERWISE IT
%READS LIKE WE SHOULD BE USING THEM <-- Done???

%\begin{figure}[!ht]
%  \centering
%  \includegraphics[width=0.35\textwidth]{background/a2-voting.png}
%  \caption{Network-wide voting in A\textsuperscript{2} over 3 nodes \cite{a2}.}
%  \label{fig:a2-voting}
%\end{figure}
%
%As can be seen in Figure \ref{fig:a2-voting}, voting is very similar to a
%one-to-many broadcast in Chaos. The coordinator proposes a given value (as a
%payload) and the flags section of each packet is used to carry information on
%the state of the vote (2 bits of information are necessary for each node:
%\texttt{0x00} represents no vote cast, \texttt{0x01} is a vote in favour,
%\texttt{0x10} is a vote against. Voting flags get merged at every slot and the
%aggregated packet is continuously broadcast until completion (i.e. all votes
%have been cast, and all nodes know about the outcome).

%%%%%%%%%%%%%%%%%%%%%%%%%%%%%%%%%%%%%%%%%%%%%%%%%%%%%%%%%%%
%Baloo makes ST programming possible...

\subsection{Baloo}

Although a number of ST communication primitives exist such as one-to-all
(Glossy) and all-to-all (Chaos), they are very difficult to program because
they rely on the low-level control of timers and radio events. 
Implementing a network-stack using ST is challenging and time-consuming.
Baloo is a middleware layer \cite{baloo} that addresses this problem. Baloo
exposes a well-defined interface to enable the run-time control of ST-primitives by
the network layer and makes it possible to create higher level abstractions using ST. 
%???SUDDEN JUMP TO SAYING WE USE IT WITHOUT JUSTIFICATION <-- It provides an abstraction??? 
It is for these reason that XPC leaverages the abstraction of ST communication provided
by Baloo in its design and implementation. It is important to note that Baloo offers a 
standardised ST layer so that various ST approaches can developed in a comparable way.
%???SO WHAT, NOT SURE WHAT YOU ARE SAYING, BE MORE CLEAR ABOUT WHAT IT DOESN'T DO, WHAT WE BRING TO THE PARTY???
%%% -> Removed for Brevity
%(see Figure \ref{fig:baloo-design}).

%Baloo allows for the implementation of a wide variety of network layer
%protocols that can use multiple ST-primitives, potentially switching them at
%runtime. The middleware is designed not to impact the time synchronization
%requirements of ST-primitives.

%With the abstraction introduced by Baloo (Figure \ref{fig:baloo-middleware}) it

In Baloo, the protocol implementation is separated from the
lower level manipulation of data packets, data transfers and timing model. The
underlying ST primitives (such as Glossy or Chaos) may be changed without
affecting the 
%???WHICH - THE USER DEFINED??? 
protocols themselves. Higher level protocol logic
can then be implemented using callback functions. 

Time Division Multiple Access (TDMA) \cite{tdma} is used by Baloo to create execution rounds
and requires a fixed execution time upper time bound for each round.  All vital
protocol information is sent to the network by a central node on the first slot
of each communication round. 

%($\S$\ref{sec:baloo-callback}).

%%% -> Removed for Brevity <-
%\begin{figure}[!ht]
%\centering
%\begin{subfigure}{.25\textwidth}
%  \centering
%  \includegraphics[width=.5\textwidth]{background/baloo-design.png}
%  \caption{Overview of Baloo's design \cite{baloo}.}
%  \label{fig:baloo-design}
%\end{subfigure}%
%\begin{subfigure}{.25\textwidth}
%  \centering
%  \includegraphics[width=.5\textwidth]{background/baloo-state.png}
%  \caption{State machine of Baloo's middleware \cite{baloo}.}
%  \label{fig:baloo-state}
%\end{subfigure}
%\caption{Baloo components: design overview and middleware state-machine}
%\end{figure}

%%% -> Removed for Brevity <-
%Baloo is a state-machine (see Figure \ref{fig:baloo-state})
%with three states: bootstrapping, running and suspended. 

%???TOO MANY PARAGRAPHS STARTING WITH THE WORD ALOO <-- Fixed??? 
Baloo is driven by control packets which are sent by a controller node at the
beginning of each round. The packets contain schedule information (i.e. how to
execute the current communication round, and when to wake up for the next
round), and configuration information (i.e. slot duration and retransmission
count). Nodes that successfully receive and decode control packets can
transmit during the subsequent allocated data slots.

While Baloo offers a much needed abstraction layer for ST, it does not offer
any services such as those required for voting. In this work we leverage Baloo 
to provide services to build atomic commit protocols. 

%% file: Sections/Design-Implementation.tex
\section{XPC and \textit{Hybrid}}
In this section we present XPC and \textit{Hybrid}. XPC is a software library that provides 
abstractions for the implementation of atomic commit protocols. Hybrid is way of using 
both Glossy and Chaos for fast and reliable flooding. The challenge here is to guarantee that the two ST primitives
are used at the correct time with the correct parameters to ensure that they 
complement each other and provide the best of both approaches. This is non-trivial because 
their timing requirements, message structures and control structures are all different.

\subsection{XPC Overview}

XPC is designed to create atomic commit protocols such as 2-phase or 3-phase
commit for WSN. It makes certain assumptions that are common to sensing
applications, like the existence of a global host, although that host does not
require more resources than the non-host nodes, 
which initiates the
protocol phases. All of the other nodes act as participants.  The XPC global
host is in charge of the atomic commit protocol's overall progress from a network wide point
of view. The other nodes either commit or abort a value specified by the global
host.

When a new phase begins the XPC global host generates a transmit schedule for the all of the
nodes in the network. It sends the schedule
in a control packet. Each phase may consist of many rounds, depending on how
many nodes respond in the first round. In a network with no interference and
good communication links, a phase may only last one round (two in XPC, explained
below) if all of the nodes
reply in that round. If some nodes do not reply, then a re-transmission round
must be scheduled to request communication from only the nodes that did not
reply in the first round.

Each schedule must be generated one round in advance. An additional
final round is scheduled to handle the potential re-transmissions. 
Retransmissions to collect lost responses from the nodes can only occur a
maximum number of times to ensure that the protocol does not wait forever. This
is he retransmission limit. 
If global host fails to hear from all of the nodes after the retransmission
limit is reached a time-out will occur and cause the protocol to abort. The
protocol can then restart and propose a new value for the network to commit.

Nodes that receive a control packet respond to the XPC global host with the requested
information based on the phase of the protocol(a \texttt{vote} for the voting phase, or
a \texttt{haveCommitted} for the commit phase). A node may miss a control packet from
the global host due to interference in the network. Nodes are allowed to timeout 
if they do not receive any information after a set period of time. This is to prevent possible
deadlock conditions caused by infinitely waiting for a message from the global
host. If a node reaches a \texttt{ABORT\_STATE} due to a timeout, it behaves as
if it had received a \texttt{DO\_ABORT} message from the global node.

%%%%%%%%%%%%%%%%%%%%%%%%%%%%%%%%%%%%%%%%%%%%%%%%%%%%%%%%%%%%%%%%%%%%%%%%%%%%%%%%%%%%%%%%%%%%%%%%%%%%

\subsection{XPC fundamentals} \label{sec:xpc-fundamentals}

\begin{figure}[!ht]
  \centering
  \includegraphics[width=.25\textwidth]{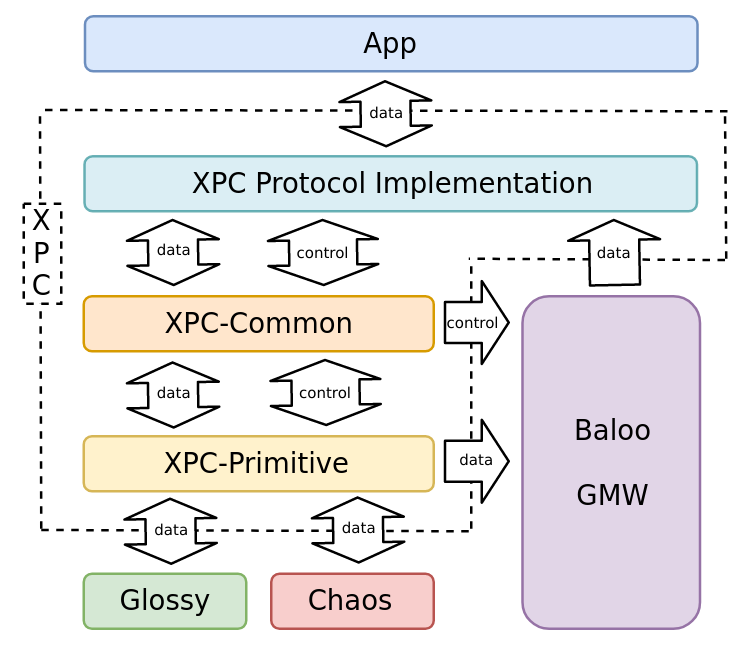}
  \caption{Layered overview of all XPC components. XPC lives alongside Baloo's
  implementation, processing all communication from the application and managing
  the commit protocol.}
  \label{fig:xpc-structure-overview}
  %\vspace{-7pt}
\end{figure}

XPC is a stack of four components (see Figure
\ref{fig:xpc-structure-overview}) to remove the application developer
from the lower ST communication layers: 

%% -> Removed for brevity. <-
\begin{enumerate}
    \item \textbf{Application}: XPC provides abstractions to implement 
    atomic commit protocols. The decision to commit or abort a transaction is
    provided as a service.  The atomic commit protocol can include the
    execution of arbitrary code. Nodes poll upon completion and execute all XPC
    code prior to the application being pre-empted. This creates a strong
    separation of application and protocol.  
    \item \textbf{Protocol implementation}: Protocols are only required to handle
    their internal logic and packet processing.
    internal state-machine transitions and packet processing.
    Network dissemination and primitive-specific tuning is handled via specific
    API calls to other layers of the XPC stack.
    \item \textbf{Common code}: Packet buffers, message
    parsing, and all re-transmission policies are packaged into a
    ``common'' section. Protocol code is simple and readable.
    \item \textbf{ST primitives}: XPC has an API that abstracts away the
    requirements of the ST primitive. Each ST primitive has completely
    different message packet structures and timing requirements. 
    With XPC, a protocol can specify which ST primitive to use to exchange
    messages for each round. 
\end{enumerate}

%%%%%%%%%%%%%%%%%%%%%%%%%%%%%%%%%%%%%%%%%%%%%%%%%%%%%%%%%%%%%%%%%%%
\subsection{Baloo Control}
\begin{figure}[!ht]
  \centering
  \includegraphics[width=.40\textwidth]{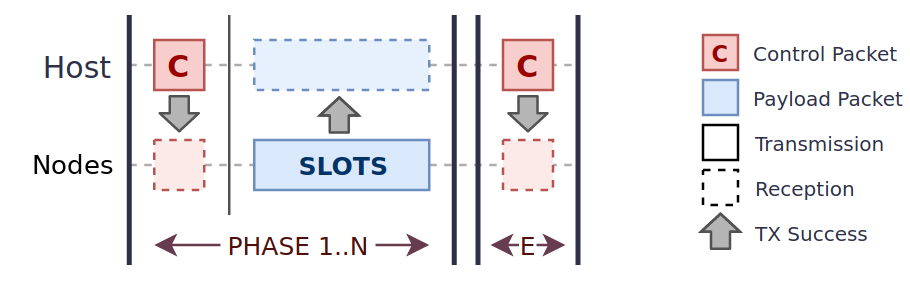}
  \caption{Example X-Phase protocol ported to Baloo's round structure. }
  \label{fig:baloo-protocol-porting}
\end{figure}

XPC uses and configures Baloo in the following ways:

\begin{enumerate}
    \item \textbf{Single Initiator}. Baloo relies on the presence of a global
    host. This node is in charge of bootstrapping the network and sending
    control packets at the beginning of each flood. With XPC, the global host
    is in charge of the protocol and the protocol state machine. 
    \item \textbf{Retransmissions for Reliability}. WSN links are very
    unreliable and packets may be lost due to interference or environmental
    conditions. To mitigate this issue XPC uses retransmissions
    to execute a phase more than once should
    there be missing replies. 
    \item \textbf{Additional Final Round}. 
    At the beginning of a Baloo round we cannot be sure whether the XPC global host will
    receive replies from all of the nodes. If it does not, XPC schedules a
    ``retransmission'' round to request information from the hosts that did not
    reply. XPC schedules a final, empty round to
    handle the potential for missing replies. In Figure
    \ref{fig:baloo-protocol-porting}, the XPC global host sends control packet $C$ during
    rounds $1$ to $N$, and expects all of the nodes to reply during their
    scheduled slots. If all of the nodes successfully reply by round $N$, a
    final empty round (denoted as $E$) is scheduled to communicate protocol
    termination. If not, another communication schedule will be sent. 
    The empty round was chosen to give the protocol the flexibility to 
    schedule or cancel retransmission rounds dynamically based on the number
    of nodes that respond.
\end{enumerate}

In order to support these adjustments, there are a number of common Baloo
configurations that will be used by all protocols, regardless of their
underlying ST primitive:
\begin{itemize}
    \item \texttt{schedule.period}. All protocols share the same length of time
    allocated for the execution of the application after a successful iteration
    of the protocol. This is configured to adapt the protocol runs to the
    needs of the top-level application.
    \item \texttt{user\_bytes}. All protocol information necessary for a given
    round is
    disseminated in a control packet. The host assigns two
    sections of the optional \texttt{user\_bytes} configuration parameter: the
    first holds the message sent by the host to all nodes in the network, the
    second holds the value currently proposed by the host.
\end{itemize} 

\noindent XPC is built alongside Baloo and uses its callback structure as seen in Figure
\ref{fig:xpc-structure-overview}. \\

%%%%%%%%%%%%%%%%%%%%%%%%%%%%%%%%%%%%%%%%%%%%%%%%%%%%%%%%%%%%%%%%%%%

\subsection{XPC Global Host and Participants}

XPC presents an API for stateless primitive manipulation. Protocols that use
XPC only have to implement their logic while using on a simplified minimal
Baloo callback structure. 

In XPC one node is the global host and all other nodes are participants.  We
present pseudo-code for an XPC initiator node (see
Algorithm \ref{alg:xpc-global-host}) and an XPC participant node (see Algorithm
\ref{alg:xpc-participant}). Within the pseudo-code we denote as parameters to
the callback functions all variables present within the global state of the
node that will be modified during the execution of the function itself. All
variables are passed by reference and updates using the $\gets$ operator modify
the internal state of the object referenced by the given variable.

\begin{algorithm}[!ht]
\caption{XPC Global Host Pseudocode}
\label{alg:xpc-global-host}
    \begin{algorithmic}[1]
      \Function{on\_round\_fin}{$\vars{state}, \vars{mess}, \vars{retr\_cnt}, \vars{reply\_num}$} \label{op:round-finish-ini}
        \State{$\vars{retr\_cnt} \gets \vars{retr\_cnt} + 1$}
        \State{$\vars{n\_slots} = \vars{XPC\_COHORT\_NODES}$ - $\vars{reply\_num}$} \label{line:nslots-set}
        \If{$\vars{retr\_cnt} >  \vars{XPC\_TIMEOUT\_RETX}$} \label{line:retr-cnt}
            \State $\vars{state} = \vars{XPC\_ABORT\_STATE}$
        \ElsIf{$\vars{reply\_num} ==  \vars{XPC\_COHORT\_NODES}$} \label{line:repl-all-init}
            \State \Call{state\_transition}{$\vars{state}$}
            \State $\vars{n\_slots} = \vars{XPC\_COHORT\_NODES}$
            \State{$\vars{retr\_cnt} \gets 0$} \label{line:repl-all-end}
        \EndIf{}
        \State \Call{prepare\_message}{$\vars{state}, \vars{mess}$} \label{line:ctrl-prep-init}
        \State Prepare the Control Packet with the $\vars{mess}$ and $\vars{n\_slots}$ \label{line:ctrl-prep-end}
      \EndFunction
      \State
      \Function{on\_slot\_post}{$\vars{state}, \vars{mess}, \vars{reply\_num}, \vars{node\_id}$}
        \If{we have not yet received a reply from $\vars{node\_id}$ while in this $\vars{state}$}
            \State \Call{process\_message}{$\vars{state}, \vars{mess}, \vars{reply\_num}$}
        \EndIf{}
      \EndFunction
  \end{algorithmic}
\end{algorithm}

The XPC global host (can be the same as the Baloo host) is in charge of the protocol's
progress and therefore determines the contents of the control packets for each
round (Algorithm \ref{alg:xpc-global-host}). Control packets must always be
generated one round in advance. The \texttt{on\_round\_finished} callback
is executed at the end of round $t$ and generates the new control packet which
will be sent to initiate round $t+1$.

When generating a control packet the global host will first determine how many nodes must
be scheduled to reply within the given round (set using the \texttt{n\_slots}
field on line \ref{line:nslots-set}). It then determines if it has
retransmitted more than the retransmission limit (line \ref{line:retr-cnt}). If
all nodes have replied during the previous round (lines
\ref{line:repl-all-init}-\ref{line:repl-all-end}) the global host
updates its state to execute the next protocol phase. It then prepares a
control packet with the correct \texttt{message} to send when it reaches 
the given state (lines \ref{line:ctrl-prep-init}-\ref{line:ctrl-prep-end}).

All other protocol logic is executed within the \texttt{on\_slot\_post}
callback. The global host processes replies from all of the nodes and
determines if the correct nodes have replied and how each reply impacts its
state machine. The protocol implemented using XPC will determine
\texttt{state\_transition(...)}, \texttt{prepare\_message(...)} and
\texttt{process\_message(...)}.

\begin{algorithm}[!ht]
\caption{XPC Participant Pseudocode}
\label{alg:xpc-participant}
    \begin{algorithmic}[1]
      \Function{on\_ctrl\_slot\_post}{$\vars{state}, \vars{ctrl}, \vars{mess}, \vars{retr\_cnt}$}
        \State{$\vars{retr\_cnt} \gets \vars{retr\_cnt} + 1$} \label{line:phase-change-1}
        \If{$\vars{retr\_cnt} >  \vars{XPC\_TIMEOUT\_RETX}$}
            \State $\vars{state} = \vars{XPC\_ABORT\_STATE}$ \label{line:phase-change-2}
        \EndIf{} 
        \If{$\vars{state} == \vars{XPC\_ABORT\_STATE}$}
              \State $\vars{mess} \gets \vars{XPC\_DO\_ABORT}$
            \EndIf{}
        \State{$\vars{prev\_state} = \vars{state}$} \label{line:phase-change-3}
        \State \Call{state\_transition}{$\vars{state}, \vars{ctrl}, \vars{mess}$} \label{line:ctrl-pack-parse}
        \If{$\vars{prev\_state} \neq \vars{state}$} \label{line:phase-change-4}
            \State{$\vars{retr\_cnt} \gets 0$} \label{line:phase-change-5}
        \EndIf{}
      \EndFunction
      \State
      \Function{on\_slot\_pre}{$\vars{state}, \vars{mess}, \vars{node\_id}$}
        \If{it is $\vars{node\_id}$'s turn to transmit}
            \State Send the $\vars{mess}$ using the correct primitive
        \EndIf{}
      \EndFunction
  \end{algorithmic}
\end{algorithm}

Participants in an XPC round (see Algorithm
\ref{alg:xpc-participant}) will parse the control packet and attempt to execute
a state transition based on the new information sent by the host (line
\ref{line:ctrl-pack-parse}). Similar to the XPC global host, all participant
nodes are allowed to timeout if the information sent by the global host does not cause
a change in their state after a set amount of retransmissions (lines
\ref{line:phase-change-1}-\ref{line:phase-change-2}, \ref{line:phase-change-3},
\ref{line:phase-change-4}-\ref{line:phase-change-5}). This is to prevent
deadlock. A global host may miss too many replies from a
participant and time-out. The current round is aborted and then a new value is
proposed. Due to interference in the network a participant may miss a control
packet containing timeout-abort information and may infinitely wait for a
specific message from the global host, stalling the whole network for multiple rounds.
If a participant reaches an \texttt{ABORT\_STATE} due to a timeout it behaves
as if it had received a \texttt{DO\_ABORT} message. The protocol that is
implemented with XPC will determine the 
\texttt{state\_transition(...)}.

%%%%%%%%%%%%%%%%%%%%%%%%%%%%%%%%%%%%%%%%%%%%%%%%%%%%%%%%%%%%%%%%%%%

\subsection{\textit{Hybrid} Synchronous Transmission Scheme}

Chaos and Glossy can not be naively combined and be expected 
to provide low latency and high reliability. We have to determine the appropriate 
time to use each primitive and what parameters they require.

For Chaos it is important to select the correct slot duration and number of
retransmissions. If the Chaos slot duration is too short, few node replies will
be received and many retransmissions may be required to receive replies from
the remaining nodes. If the initial Chaos slot duration is too long, time may
be wasted waiting for replies from nodes. 

For Glossy it is important to use it for as few nodes as possible, as each node
requires a separate flood. 

The Chaos slot duration also has to be bounded in order to fit the TDMA slots
of Baloo. The original implementation of Chaos allowed the nodes to communicate
until the global host was certain that it had received all of the nodes'
information. In section \ref{sec:xpc-chaos} we determined the best Chaos slot 
duration experimentally for the FlockLab testbed.   
We leave the question of dynamic slot sizes that adapt to network conditions for
future work.

\begin{figure}[!t]
\vspace{-10pt}
  \centering
  \includegraphics[width=.5\textwidth]{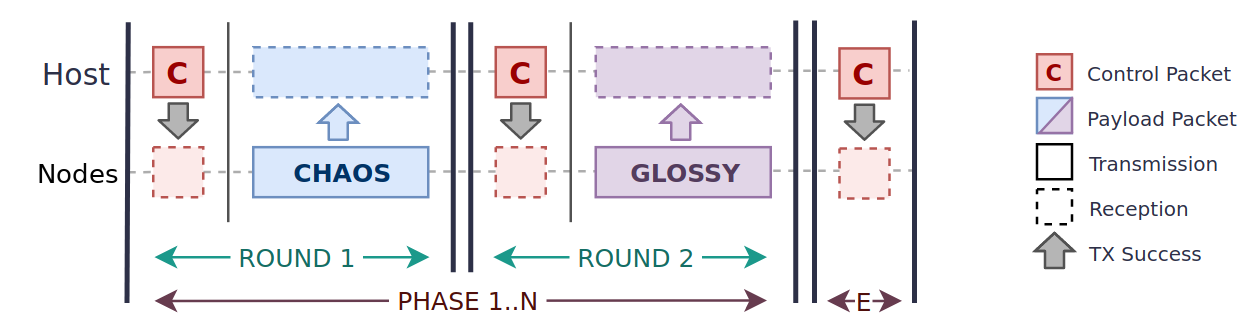}
  \caption{Overview of \textit{Hybrid} execution across multiple phases. Each phase
  starts with a Chaos dissemination round,  followed by a variable number of
  Glossy floods.}
  \label{fig:xpc-hybrid-protocol}
\end{figure}

\textit{Hybrid} organises the use of both Chaos and Glossy. 
The first
transmission of each phase is executed using a Chaos flood with a slot duration
selected in order to reach as many node as possible (see Figure
\ref{fig:xpc-hybrid-protocol}). If all of the nodes have not responded, then
Glossy is then used for reliable retransmissions to collect the remaining
information. The use of Chaos first gives us a fast (but unreliable) way to get
most of the network. Glossy is then used to get the remaining nodes reliably.
Up to a maximum of $\texttt{XPC\_TIMEOUT\_RETRANSMISSIONS} - 1$ Glossy floods
may be scheduled, after which, if still no reply is heard from every node, the
protocol times-out.

\textit{Hybrid} has to manage the change of packet types used for both Chaos
and Glossy, and inform the nodes of the change in ST primitive. It achieves
this using the control packet of the XPC global host. 

In the remainder of this paper we use XPC to compare Glossy, Chaos, and \textit{Hybrid}
used in atomic commit protocols. \textit{Hybrid} could be used, in principle, for any
application that requires flooding. At the moment, our only implementation of \textit{Hybrid}
is with XPC.

We used XPC to create two reference atomic commit protocols for the purposes of
evaluation. We implemented 2-phase commit \cite{twopc} and 3-phase 
commit\cite{threepc} using XPC with
Glossy only, with Chaos only, and with \textit{Hybrid}. These atomic commit protocols 
are representative of the type of commit protocols that WSN would use to agree
upon new sensing parameters or other system configuration.

%% file: Sections/Evaluation.tex
\section{Evaluation} \label{sec:xpc-top-hybrid}

In this section we report experimental results comparing the
latency and reliability of Glossy, Chaos and our Hybrid approach. 
We were not able to evaluate against other ST approaches like
Synchotron\cite{al2017network} and Crystal \cite{trobinger2019competition} 
because we were unable to get the code to work reliably, 
a well documented issue highlighted by other researchers, 
%???MENTION AMI'S WORK <-- Added, but may be cut due to space????
however we compare our figures with published Synchotron values running in the same testbed. 
%???IS THIS CORRECT CHECK ME <-- Yes???
%???WHY WAS SYNCATRON NOT USED <-- We could not get them to work.???

We used XPC to create two reference atomic commit protocols for the purposes of
evaluation. We implemented 2-phase commit(2PC)\cite{twopc} and 3-phase 
commit(3PC)\cite{threepc} using XPC with
Glossy only, with Chaos only, and with \textit{Hybrid}. These atomic commit protocols 
are representative of the type of commit protocols that WSN would use to agree
upon new sensing parameters or systems configurations.
%Each is used
%as the Synchronous Transmission (ST) layer for the two-phase commit(2PC) and three-phase
%commit(3PC) protocols implemented using XPC.
We evaluate the agreement outcome and the latency of each of the atomic commit 
protocol with each ST primitive. The we experimentally evaluate the robustness 
of the atomic commit protocols with different degrees of radio interference.
Finally, we compare our latency results to the reported results of
A\textsuperscript{2}/Synchrotron which uses Chaos\cite{al2017network}.  

Our analysis was performed using 22 and 27 TelosB nodes(depending on
availability at the time) on the FlockLab testbed \cite{flocklab}. FlockLab has resources 
to monitor the low-level execution of the nodes that is mentioned in 
section \ref{sec:xpc-chaos} and has been one the main testbeds used in ST research. 
Given the inherent scale limitations of both Glossy and Chaos (scale is an open 
problem in ST research), we believe that the 
FlockLab testbed provides an adequate network size and network density for our evaluation. 
%???OK THIS DOES NOT JUSTIFY THE SMALL SCALE WELL -- SAY SOMETHING LIKE FLOCKLAB PROVIDES 
%FACILITIES
%TO TEST MANY DIFFERING RADIO CONDITIONS AND THEREFORE IS USED IN THESE EXPERIEMNTS, IT HAS BEEN 
%USED IN THE PAST FOR OTERH SYNC PROTOCOL EXPERIMENTS. HOWEVER ALL SCALE LIMITATION ARE INHERITED 
%FROM THE UNDERPINNING GLOSSY/CHAOS PROTOCOLS AND NOT OUR XPC OR HYBRID IMPLEMENTATIONS AS THEY
%MAKE
%DECISIONS AGNOSTIC OF THE SIZE OF THE NETWORK --- MAKE SURE WHAT YOU SAY IS TRUE???
% We inherit scale limitations from Glossy and Chaos.
%???SAY SOMETHING ABOUT SCALE HERE - DO WE WORRY ABOUT THAT NOT BEING LARGE - SAY WHY WE DON'T WORRY <-- Neither can scale.???

%%%%%%%%%%%%%%%%%%%%%%%%%%%%%%%%%%%%%%%%%%%%%%%%
\subsection{XPC using Glossy} \label{sec:xpc-glossy}

\begin{figure}[!ht]
\centering
\begin{subfigure}{.25\textwidth}
  \centering
  \adjustbox{trim={0} {0} {0.5\width} {0},clip}%
  %\adjustbox{trim={0} {0.5\width},clip}%
    {\includegraphics[width=2\textwidth]{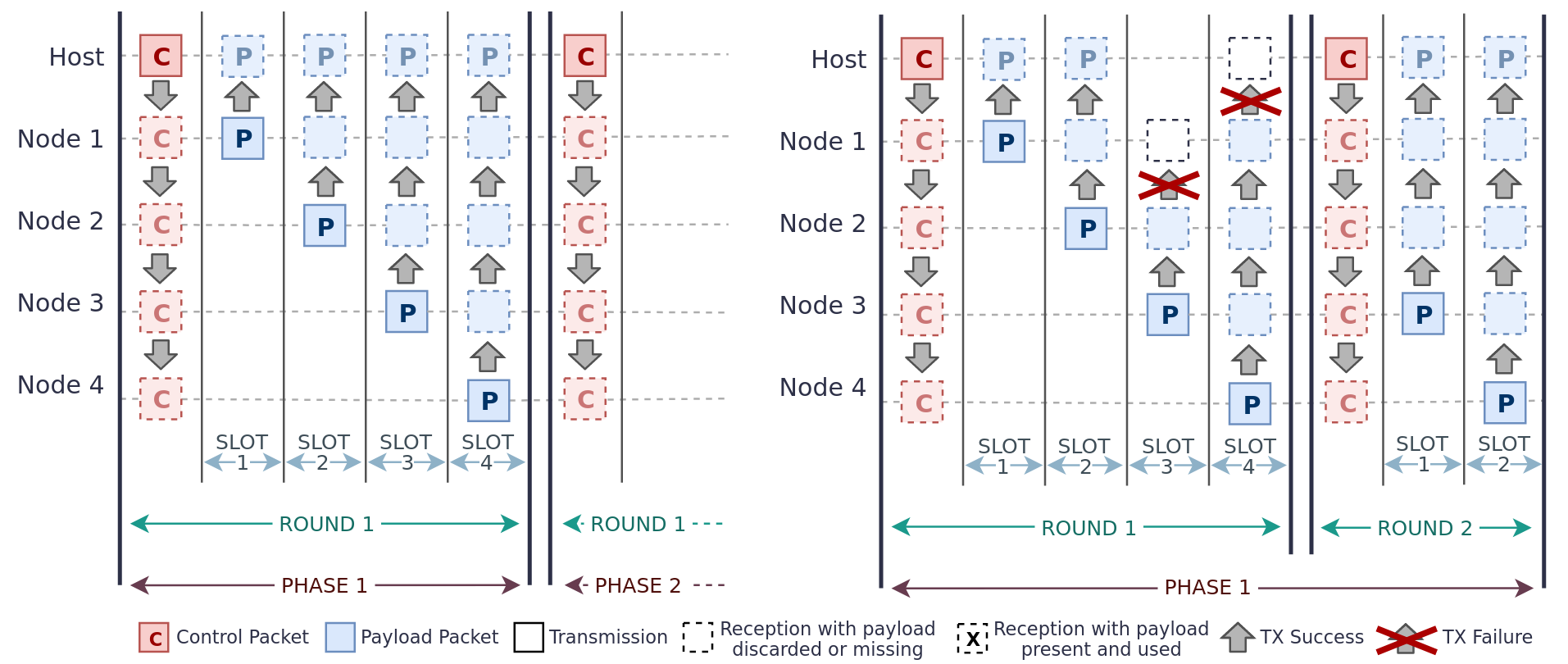}}
  \caption{Glossy round execution}
  \label{fig:baloo-glossy-protocol}
\end{subfigure}%
\begin{subfigure}{.25\textwidth}
  \centering
  \adjustbox{trim={0.5\width} {0} {0} {0},clip}%
  %\adjustbox{trim={0.5\width} {0},clip}%
    {\includegraphics[width=2\textwidth]{voting/baloo-glossy-protocol.png}}
  \caption{Glossy retransmission round}
  \label{fig:baloo-glossy-retransmission}
\end{subfigure}
\caption{Execution of Glossy rounds with XPC. When nodes do not reply in a
given slot they are scheduled to retransmit in the subsequent round during the
same phase.}
\label{fig:xpc-glossy}
\end{figure}

XPC with Glossy uses a time-sliced data dissemination
approach. Given a network of $k$ nodes (where one is the XPC global host), each
round a \texttt{schedule.n\_slots} field is set to $k - 1$. All
nodes, except the global host, communicate in a given
\texttt{schedule.slot} (see Figure \ref{fig:baloo-glossy-protocol}). The nodes
receive round information from the global host within the control packet.
Nodes reply to the host by broadcasting during their scheduled
slot. The \texttt{payload} exchanged during each round contains the messages
sent by each node as a reply to the host. When XPC uses Glossy, each reply is one byte.

The Glossy-based protocols keep track of nodes that do not reply in their
scheduled slots and schedule them to re-transmit in the next round. The
re-transmission round will have a length equal to the number of nodes that did
not reply. In Figure \ref{fig:baloo-glossy-retransmission}, node $3$ and node
$4$ did not successfully send their reply to the host node during the first
round. The re-transmission round (i.e. round 2) contains slots for only those two
nodes' \texttt{schedule.slot}.

%%%%%%%%%%%%%%%%%%%%%%%%%%%%%%%%%%%%%%%%%%%%%%%%%%%%%%%%%%%%%%%%%%%%%%%%%%%%%%%%%%%%%%%%%%%%%%%%%%%%
\subsubsection{Two-Phase Commit with Glossy}
\begin{figure*}[!ht]
\centering
\vspace{-10pt}
\begin{subfigure}{.20\textwidth}
  \centering
  \includegraphics[width=\textwidth]{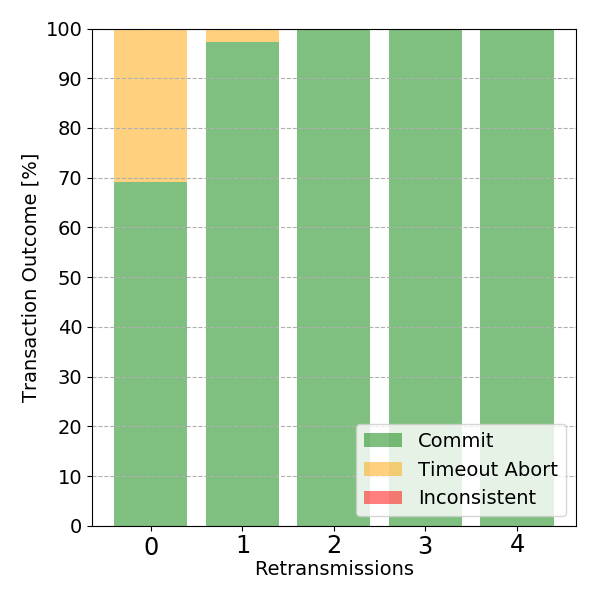}
  \caption{2PC-Glossy transaction outcome}
  \label{fig:2pc-glossy-outcome}
\end{subfigure}%
\begin{subfigure}{.20\textwidth}
  \centering
  \includegraphics[width=\textwidth]{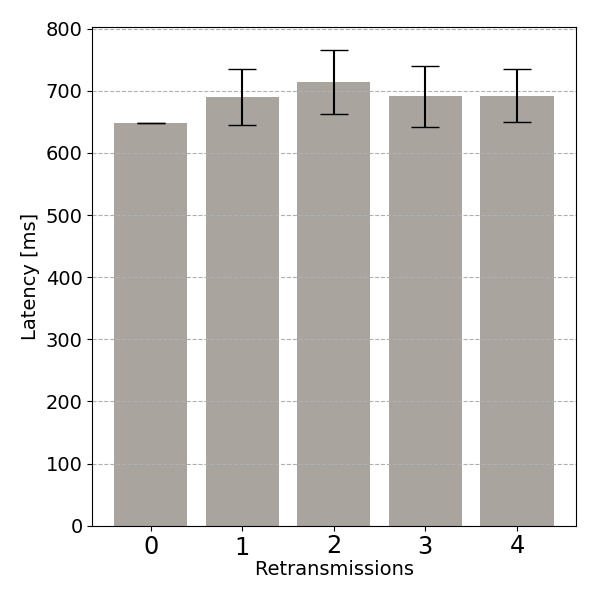}
  \caption{2PC-Glossy retransmission latency}
  \label{fig:2pc-glossy-latency}
\end{subfigure}
\begin{subfigure}{.20\textwidth}
  \centering
  \includegraphics[width=\textwidth]{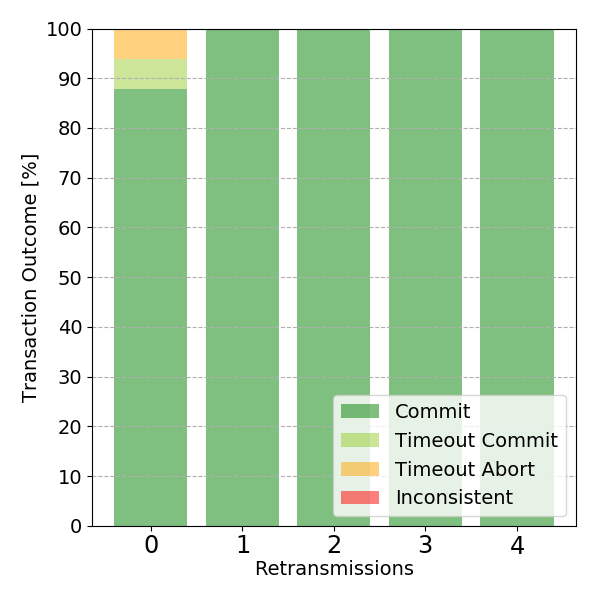}
  \caption{3PC-Glossy transaction outcome}
  \label{fig:3pc-glossy-outcome}
\end{subfigure}%
\begin{subfigure}{.20\textwidth}
  \centering
  \includegraphics[width=\textwidth]{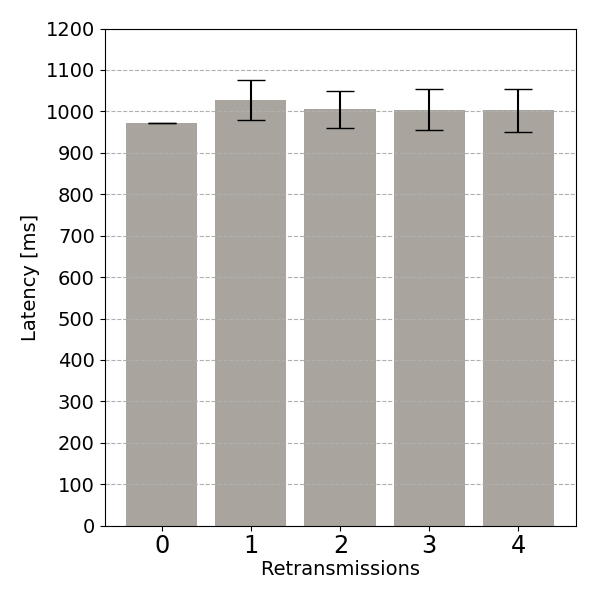}
  \caption{3PC-Glossy retransmission latency}
  \label{fig:3pc-glossy-latency}
\end{subfigure}
\caption{Evaluation of transaction outcome and latency for XPC 2PC-Glossy and XPC 3PC-Glossy in FlockLab.}
\label{fig:3pc-glossy-overview}
%\vspace{-7pt}
\end{figure*}

%Two-Phase Commit is a simple protocol which benefits greatly from a time-sliced
%reliable dissemination primitive such as Glossy's. 
Our first set of results show that 
2PC is unable to reliably reach all nodes in one round while the introduction of
re-transmission rounds greatly boosts the overall reliability (Figure
\ref{fig:2pc-glossy-outcome}).  Note that there
are very few timeout aborts when retransmissions are used.

The results in Figure \ref{fig:2pc-glossy-latency} show that retransmissions do
not significantly increase the latency of the protocol. With no retransmissions
the protocol reaches a timeout abort in approximately $30$\% of the runs. 

%%%%%%%%%%%%%%%%%%%%%%%%%%%%%%%%%%%%%%%%%%%%%%%%%%%%%%%%%%%%%%%%%%%%%%%%%%%%%%%%%%%%%%%%%%%%%%%%%%%%
\subsubsection{Three-Phase Commit with Glossy}

The results in Figure \ref{fig:3pc-glossy-outcome} show that ``timeout
commits'' do not change the reliability of 3PC-Glossy when compared to
2PC-Glossy (Figure \ref{fig:2pc-glossy-outcome}). The most significant affect
can be seen in latency (Figure \ref{fig:3pc-glossy-latency}). 3PC has one more
phase, and a higher latency when compared to 2PC. 

%\begin{figure*}[!ht]
%\centering
%\begin{subfigure}{.5\textwidth}
%  \centering
%  \includegraphics[width=.5\textwidth]{voting/3pc-transaction-outcome-retransmissions.png}
%  \caption{3PC-Glossy transaction outcome}
%  \label{fig:3pc-glossy-outcome}
%\end{subfigure}%
%\begin{subfigure}{.5\textwidth}
%  \centering
%  \includegraphics[width=.5\textwidth]{voting/3pc-latency-retransmissions.png}
%  \caption{3PC-Glossy retransmission latency}
%  \label{fig:3pc-glossy-latency}
%\end{subfigure}
%\caption{Evaluation of transaction outcome and latency for 3PC-Glossy in FlockLab.}
%\label{fig:3pc-glossy-overview}
%%\vspace{-7pt}
%\end{figure*}

%%%%%%%%%%%%%%%%%%%%%%%%%%%%%%%%%%%%%%%%%%%%%%%%%%%%%%%%%%%%%%%%%%%%%%%%%%%%%%%%%%%%%%%%%%%%%%%%%%%%
%%%%%%%%%%%%%%%%%%%%%%%%%%%%%%%%%%%%%%%%%%%%%%%%%%%%%%%%%%%%%%%%%%%%%%%%%%%%%%%%%%%%%%%%%%%%%%%%%%%%
%%%%%%%%%%%%%%%%%%%%%%%%%%%%%%%%%%%%%%%%%%%%%%%%%%%%%%%%%%%%%%%%%%%%%%%%%%%%%%%%%%%%%%%%%%%%%%%%%%%%
\subsection{XPC using Chaos} \label{sec:xpc-chaos}

%Chaos has very different timing requirements to Glossy requiring XPC
%to match the precisely timed and synchronous mechanisms provided by Baloo.
%Chaos is an all-to-all communication primitive. It terminates dissemination rounds
%when it has received a reply from all nodes. XPC has to provide a fixed
%execution time upper bound for each round in order to be used with Baloo.
%XPC modifies the \texttt{schedule.period} section of the control packet and
%sets it to a mutable value. This allows enough time for the nodes to
%reply to the Chaos initiator.  Missed nodes are reached via retransmissions. 
%???REPEATED FROM BEFORE --> Removed ???

\begin{figure}[!ht]
    \centering
    \includegraphics[width=.4\textwidth]{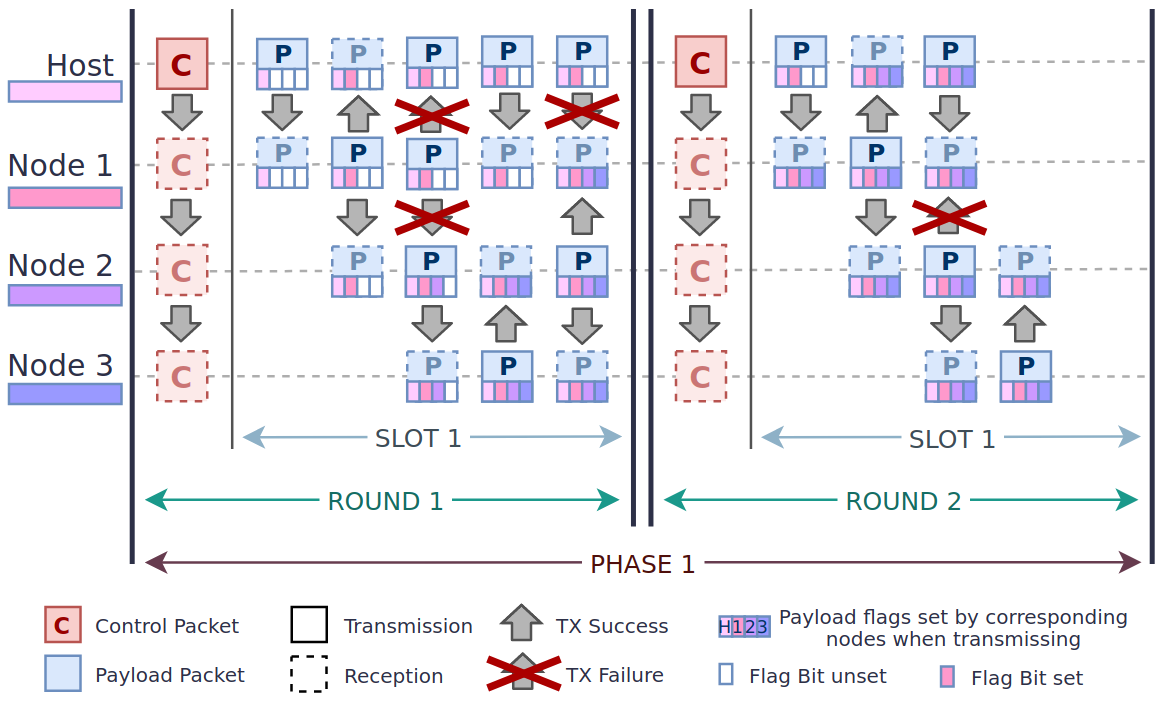}
    \caption{Execution of an XPC Chaos round with 1 re-transmission. As nodes
    aggregate their vote into the payload they set their bits into the packet's
    flags bit-field.}
    \label{fig:baloo-chaos-retransmission}
    %\vspace{-7pt}
\end{figure}

Compared to Glossy, Chaos prioritises latency over reliability. As can be
seen in Figure \ref{fig:baloo-chaos-retransmission} Chaos floods occur in a
``best-effort'' fashion. There is no certainty that a round will be long enough
for communication to reach all of nodes in the network and aggregate their replies. 
%???FEELS REPETITIVE TOO???

% as12015 C4.01
% > "Chaos floods are hard to time bound, they last until all of the nodes cease to receive new information from their neighbours"
% That is what normally occurs in A2 Synchrotron, yet by using Baloo's implementation of Chaos we are forced to specify strong timing bounds for Chaos to execute in. If Chaos overruns its maximum allocation time (regardless of completion) it gets terminated.
% So would say "Chaos floods are hard to time bound, as they were originally designed to last until ..."

The slot duration is the most important parameter when using Chaos with 2PC and
3PC. Chaos floods are hard to time bound, they were originally designed to last
until all of the nodes cease to receive new information from their neighbours.
With XPC the challenge is to bound the maximum communication time of a Chaos flood, the
slot duration.  
%Two problems can occur if the communication slot duration is
%poorly chosen.  
%
%\begin{itemize}
%    \item \textbf{Under-allocation}. Chaos slots which are too small might
%    not allow the host's message to be received by the whole network since there may
%    not be enough time for the information to propagate to all of the nodes in a
%    network with a large depth.
%    \item \textbf{Over-allocation}. A longer slot duration will increase the
%    latency. 
%\end{itemize}
%???HAS THE ABOVE NOT ALREADY BEEN DISCUSSED
Another challenge is that Chaos is unable to directly target a specific node in
the network when requesting retransmissions due to missing replies. It is also
unaware of how many iterations will be required for the missing
packets to reach the host.  XPC schedules a new round of the same length using
the \texttt{payload} of the previous round (Figure
\ref{fig:baloo-chaos-retransmission}).  Communication ceases when no node sees
new information in the packets being broadcast (as seen in Figure
\ref{fig:baloo-chaos-retransmission}). The Chaos flood is resumed exactly from
where it was last stopped by XPC, and it is given extra time so that it will
terminate.

In our experiments with 2PC-Chaos and 3PC-Chaos we explore two
parameters: the number of retransmissions and the Chaos slot duration. The
results in Figure \ref{fig:2pc-chaos-outcome} validated our 
assumptions: a longer slot duration (i.e. 100ms or 200ms) reliably achieve near 100\%
reliability, very similarly to XPC using only Glossy.
%%%%%%%%%%%%%%%%%%%%%%%%%%%%%%%%%%%%%%%%%%%%%%%%%%%%%%%%%%%%%%%%%%%%%%%%%%%%%%%%%%%%%%%%%%%%%%%%%%%%
\subsubsection{Two-Phase Commit with Chaos}

\begin{figure*}[!ht]
\centering
\begin{subfigure}{.25\textwidth}
  \centering
  \includegraphics[scale=.20]{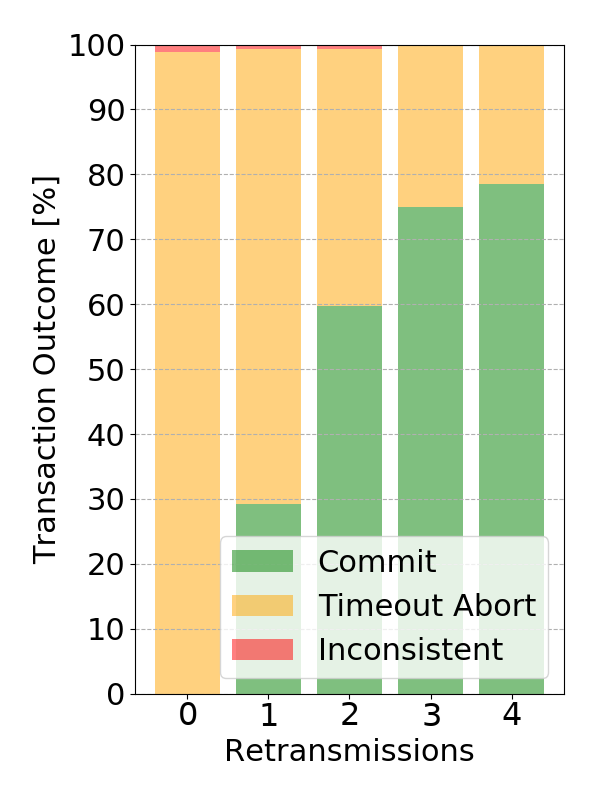}
  \caption{25ms slots}
  \label{fig:2pc-chaos-outcome-25ms}
\end{subfigure}%
\begin{subfigure}{.25\textwidth}
  \centering
  \includegraphics[scale=.20]{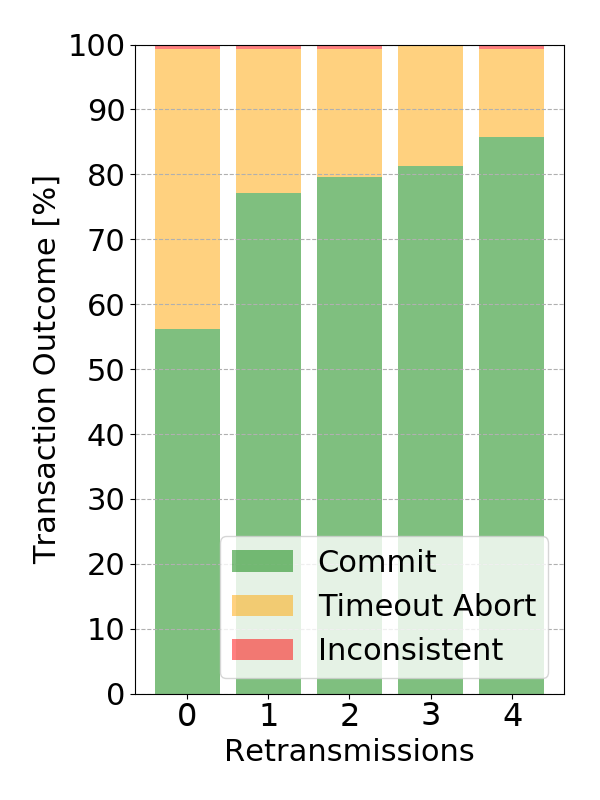}
  \caption{50ms slots}
  \label{fig:2pc-chaos-outcome-50ms}
\end{subfigure}%
\begin{subfigure}{.25\textwidth}
  \centering
  \includegraphics[scale=.20]{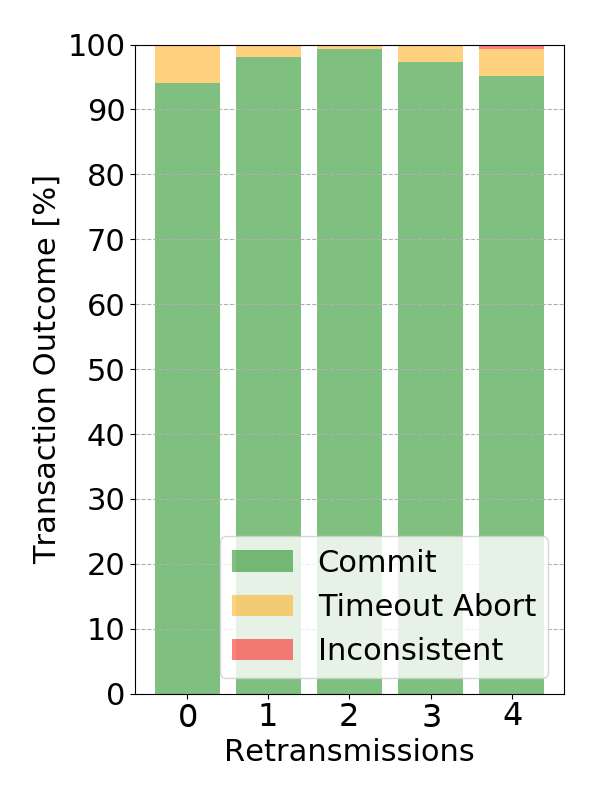}
  \caption{100ms slots}
  \label{fig:2pc-chaos-outcome-100ms}
\end{subfigure}%
\begin{subfigure}{.25\textwidth}
  \centering
  \includegraphics[scale=.20]{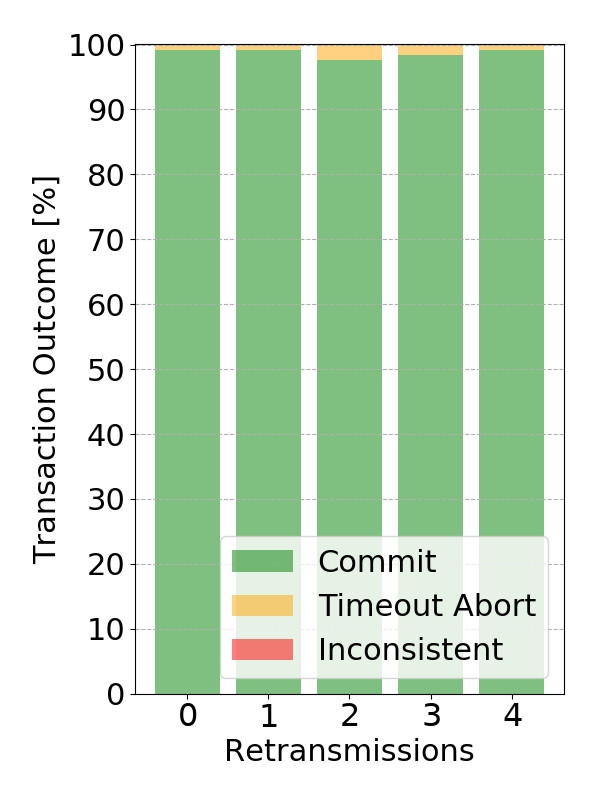}
  \caption{200ms slots}
  \label{fig:2pc-chaos-outcome-200ms}
\end{subfigure}
\caption{Agreement outcome of XPC 2PC-Chaos with varying slot duration in FlockLab.}
\label{fig:2pc-chaos-outcome}
\end{figure*}

\begin{figure*}[!ht]
\centering
\vspace{-10pt}
\begin{subfigure}{.25\textwidth}
  \centering
  \includegraphics[scale=.20]{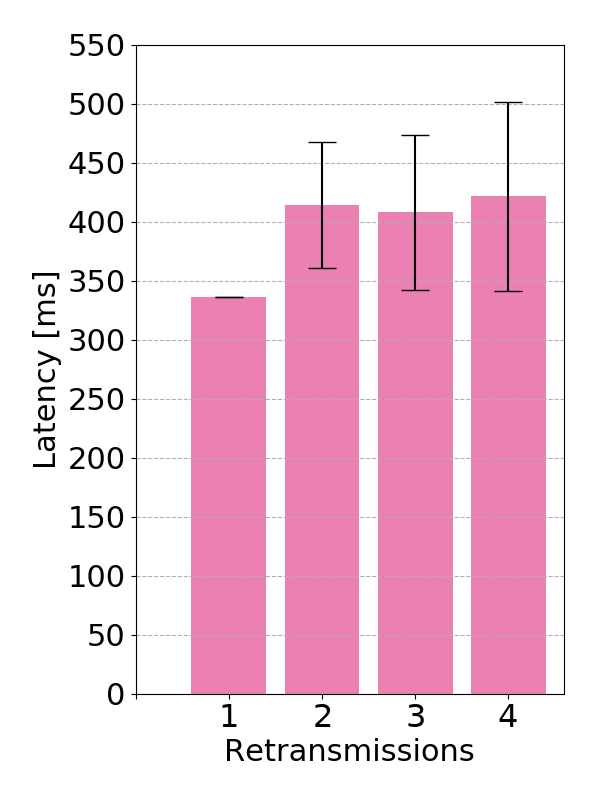}
  \caption{25ms slots}
  \label{fig:2pc-chaos-latency-25ms}
\end{subfigure}%
\begin{subfigure}{.25\textwidth}
  \centering
  \includegraphics[scale=.20]{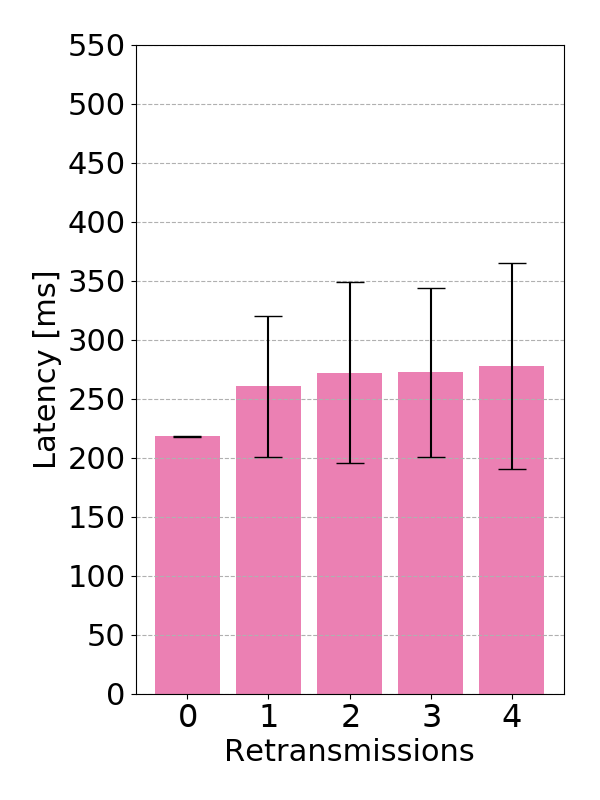}
  \caption{50ms slots}
  \label{fig:2pc-chaos-latency-50ms}
\end{subfigure}%
\begin{subfigure}{.25\textwidth}
  \centering
  \includegraphics[scale=.20]{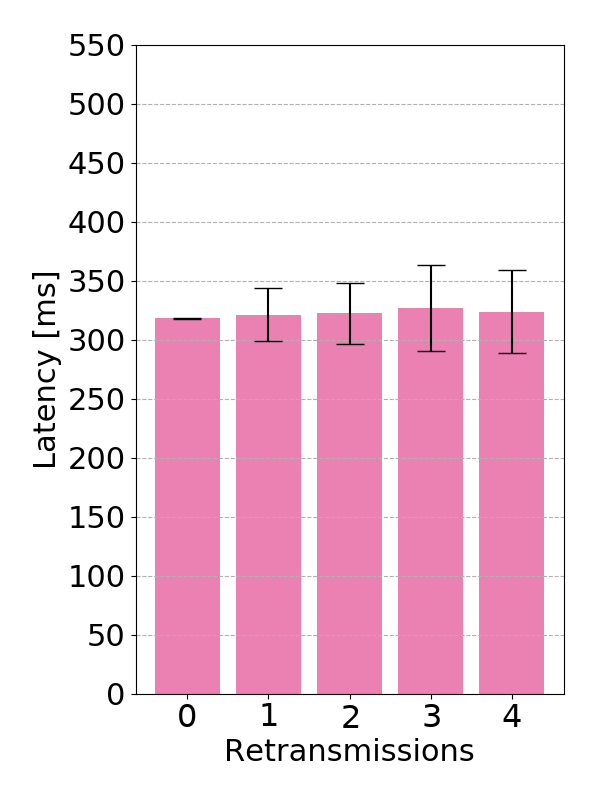}
  \caption{100ms slots}
  \label{fig:2pc-chaos-latency-100ms}
\end{subfigure}%
\begin{subfigure}{.25\textwidth}
  \centering
  \includegraphics[scale=.20]{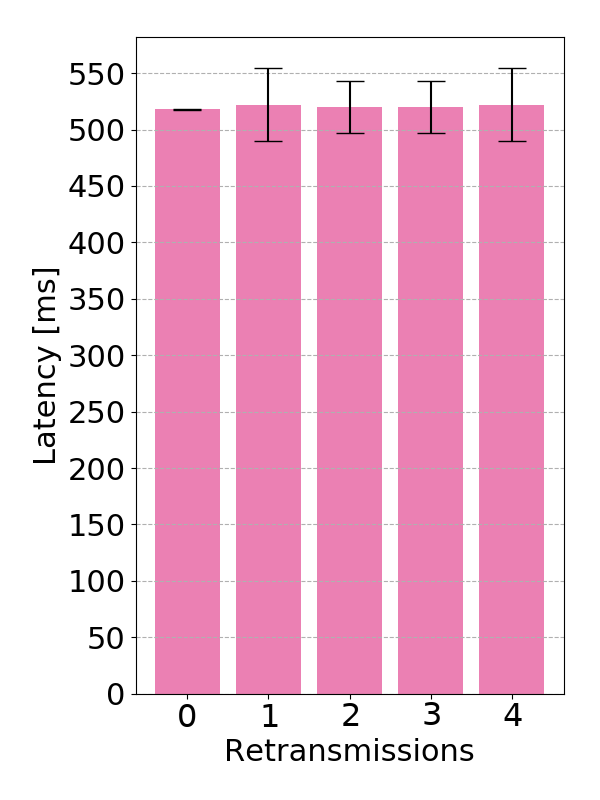}
  \caption{200ms slots}
  \label{fig:2pc-chaos-latency-200ms}
\end{subfigure}
\caption{Latency of XPC 2PC-Chaos with varying slot duration in FlockLab.}
\label{fig:2pc-chaos-latency}
\end{figure*}

The results show very poor reliability for 25ms slots (Figure
\ref{fig:2pc-chaos-outcome-25ms}) with improved results for 50ms
slots (Figure \ref{fig:2pc-chaos-outcome-50ms}). For latency, the results  
show that 25ms slots (Figure \ref{fig:2pc-chaos-latency-25ms}) have higher latency
than 50ms slots (Figure \ref{fig:2pc-chaos-latency-50ms}).

%The cause for this unreliability can be seen in the FlockLab GPIO (General Purpose Input/Output 
%pins, in this case used as LED lights) traces (Figures \ref{fig:2pc-chaos-25ms-flocklab}
%and \ref{fig:2pc-chaos-50ms-flocklab}). On the FlockLab testbed the
%red LED is set when the radio is turned on, the purple LED is set upon the 
%reception of a radio message, and the yellow LED is set when a radio message
%is broadcast. The setting of an LED has little time overhead when compared to writing output
%to serial. Flocklab offers a web visualiser for GPIO outputs. All of the logs have matching timestamps across the various
%nodes in the network.

We analysed the cause for this unreliability using FlockLab GPIO (General Purpose Input/Output 
pins, in this case used as LED lights) traces. On the FlockLab testbed the
red LED is set when the radio is turned on, the purple LED is set upon the 
reception of a radio message, and the yellow LED is set when a radio message
is broadcast. The setting of an LED has little time overhead when compared to writing output
to serial. Flocklab offers a web visualiser for GPIO outputs. All of the logs have matching timestamps across the various
nodes in the network.

% as12015 C4.02
% > "It can be seen enlarged in Figure \ref{fig:2pc-chaos-25ms-zoom}"
% So technically the 'enlarged' images are just an idealistic overview of what is going on inside Flocklab. They are not scientifically enlarged views, more rather manually drawn zoomed illustrations of what is condensed in a very tight and small pixel area. Yeah a fictional zoomed up drawing that contains the same data, but not necessarily an enlarged version where all components have a 1:1 mapping

%A GPIO pin trace of 2PC-Chaos with 25ms slots and 1 retransmission can be seen
%in Figure \ref{fig:2pc-chaos-25ms-gpio}. We show 1 retransmission because we
%are never able to commit without retransmissions (Figure
%\ref{fig:2pc-chaos-outcome-25ms}). In the image, each column is a separate run
%of the protocol. Of the 4 protocol executions present in the image only the
%final was able to commit. The final execution is represented in Figure
%\ref{fig:2pc-chaos-25ms-zoom}. All other executions ended up timing out and
%aborting due to missing replies.

\begin{figure*}[!ht]
\centering
\begin{subfigure}{.45\textwidth}
  \includegraphics[width=.6\textwidth]{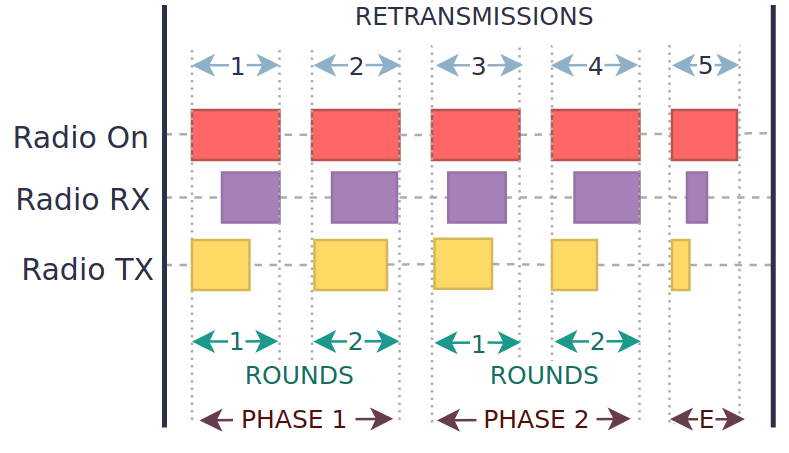}
  \caption{GPIO trace for 2PC-Chaos with 25ms slots.}
  \label{fig:2pc-chaos-25ms-zoom}
\end{subfigure}
\begin{subfigure}{.45\textwidth}
  \includegraphics[width=.6\textwidth]{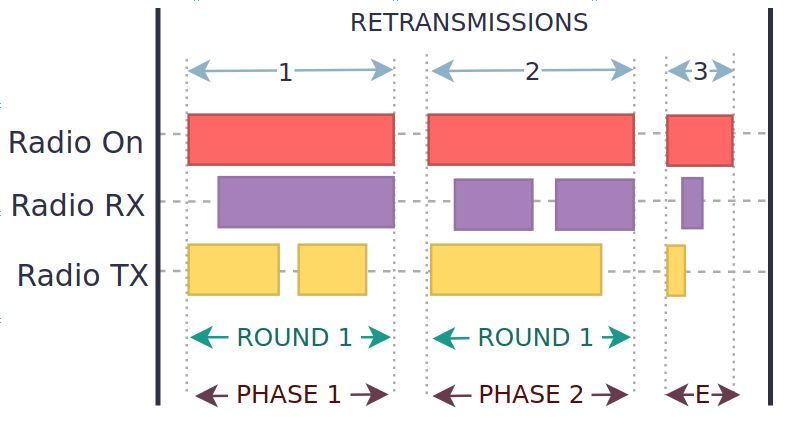}
  \caption{GPIO trace for 2PC-Chaos with 50ms slots.}
  \label{fig:2pc-chaos-50ms-zoom}
\end{subfigure}
\caption{Representations of FlockLab Radio LED GPIO tracing for 2PC-Chaos with 25ms
and 50ms slots.}
\label{fig:2pc-chaos-GPIO}
\end{figure*}

We analysed a GPIO pin trace of 2PC-Chaos with 25ms slots and 1 retransmission
because it was unable to commit without retransmissions (Figure
\ref{fig:2pc-chaos-outcome-25ms}). Most of the executions present in the GPIO
traces aborted with a time-out due to missing replies. Figure
\ref{fig:2pc-chaos-25ms-zoom} a representation of a successful commit that
required 5 retransmissions: 2 retransmissions for Phase 1, 2 retransmissions
for Phase 2, and a final re-transmission to communicate the end of the 2PC
round. The protocol uses 2 rounds for each communication phase. This is an
indication of a communication slot that is too short in duration. Not all nodes
are reached, and multiple retransmissions are required.

%The FlockLab GPIO trace for a 50ms slot duration (see Figure
%\ref{fig:2pc-chaos-50ms-gpio}) is very different. Once again we see 4 protocol
%executions. There are 3  successful commit runs, and the final execution is a
%timeout abort run.  The close-up of a successful commit execution (Figure
%\ref{fig:2pc-chaos-50ms-zoom}) shows that Chaos executes 3 rounds before
%committing. 

The FlockLab GPIO trace for a 50ms slot duration was very different. Most of
the executions in the traces were successful. The Figure
\ref{fig:2pc-chaos-50ms-zoom}) shows a representation of a successful commit
execution with 3 rounds. 

The issue is that Chaos needs gap times for its callback functions. 
If the slot duration is short, the next round occurs before the 
end of the gap times. We see that a 100ms slot duration has a higher reliability 
and slightly lower latency than any number of retransmissions with 25ms slots.

The results for 100ms and 200ms slots in Figure \ref{fig:2pc-chaos-latency}
show that 2PC-Chaos can be reliable and have a lower latency than 2PC-Glossy.
Above 95\%, reliability is achieved with 100ms long Chaos floods and 325ms
latency. Close to 100\% reliability can obtained with 200ms floods with a
latency of 525ms, around 40\% less than 2PC-Glossy when run on FlockLab. With
longer Chaos flood durations the number of retransmission numbers does not
greatly increase the overall latency. The protocol has a high reliability from
the first transmission slot.

%\begin{figure*}[!ht]
%\centering
%\begin{subfigure}{.5\textwidth}
%  \centering
%  \includegraphics[width=.6\textwidth]{voting/2pc-chaos-25ms-flocklab-gpio.png}
%  \caption{Flocklab Radio LED tracing}
%  \label{fig:2pc-chaos-25ms-gpio}
%\end{subfigure}%
%\begin{subfigure}{.5\textwidth}
%  \centering
%  \includegraphics[width=.6\textwidth]{voting/2pc-chaos-25ms-flocklab-zoom.png}
%  \caption{Breakdown of Chaos retransmissions}
%  \label{fig:2pc-chaos-25ms-zoom}
%\end{subfigure}
%\caption{Breakdown of FlockLab Radio LED tracing for 2PC-Chaos with 25ms slots.}
%\label{fig:2pc-chaos-25ms-flocklab}
%\end{figure*}
%
%\begin{figure*}[!ht]
%\centering
%\begin{subfigure}{.5\textwidth}
%  \centering
%  \includegraphics[width=.6\textwidth]{voting/2pc-chaos-50ms-flocklab-gpio.png}
%  \caption{Flocklab Radio LED tracing}
%  \label{fig:2pc-chaos-50ms-gpio}
%\end{subfigure}%
%\begin{subfigure}{.5\textwidth}
%  \centering
%  \includegraphics[width=.6\textwidth]{voting/2pc-chaos-50ms-flocklab-zoom.png}
%  \caption{Breakdown of Chaos retransmissions}
%  \label{fig:2pc-chaos-50ms-zoom}
%\end{subfigure}
%\caption{Breakdown of FlockLab Radio LED tracing for 2PC-Chaos with 50ms slots.}
%\label{fig:2pc-chaos-50ms-flocklab}
%\end{figure*}

%%%%%%%%%%%%%%%%%%%%%%%%%%%%%%%%%%%%%%%%%%%%%%%%%%%%%%%%%%%%%%%%%%%%%%%%%%%%%%%%%%%%%%%%%%%%%%%%%%%%
\subsubsection{Three-Phase Commit with Chaos}

\begin{figure*}[!ht]
\vspace{-10pt}
\centering
\begin{subfigure}{.25\textwidth}
  \centering
  \includegraphics[scale=.20]{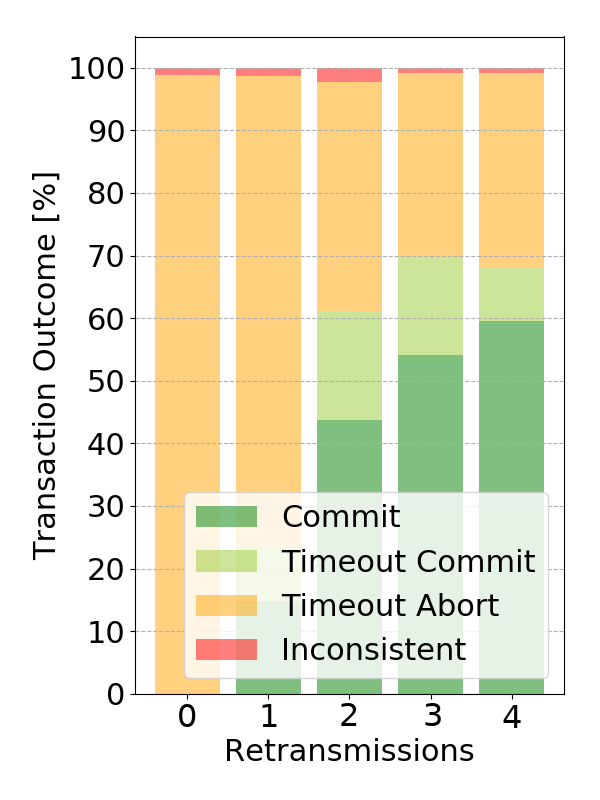}
  \caption{25ms slots}
  \label{fig:3pc-chaos-outcome-25ms}
\end{subfigure}%
\begin{subfigure}{.25\textwidth}
  \centering
  \includegraphics[scale=.20]{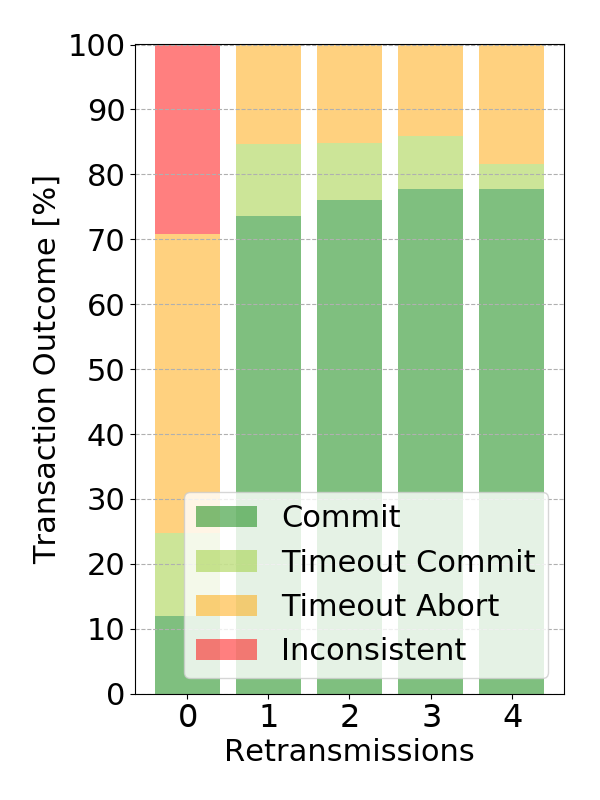}
  \caption{50ms slots}
  \label{fig:3pc-chaos-outcome-50ms}
\end{subfigure}%
\begin{subfigure}{.25\textwidth}
  \centering
  \includegraphics[scale=.20]{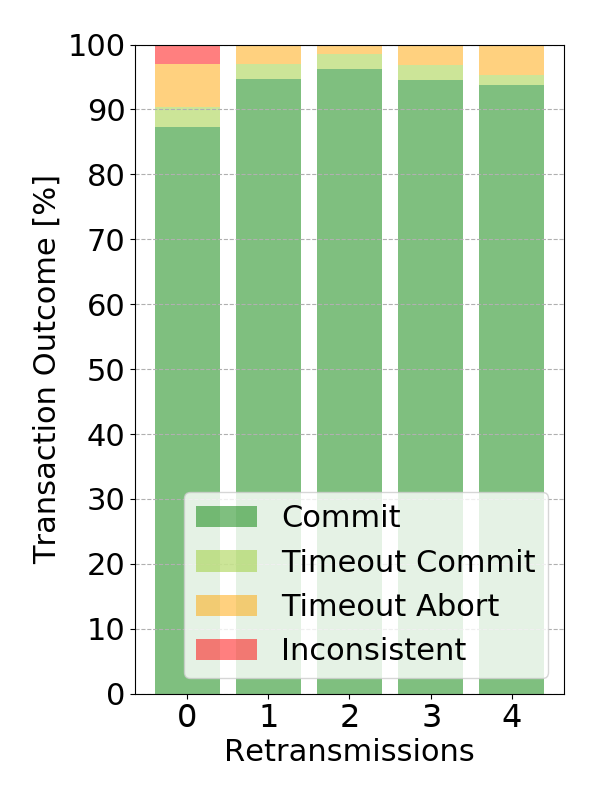}
  \caption{100ms slots}
  \label{fig:3pc-chaos-outcome-100ms}
\end{subfigure}%
\begin{subfigure}{.25\textwidth}
  \centering
  \includegraphics[scale=.20]{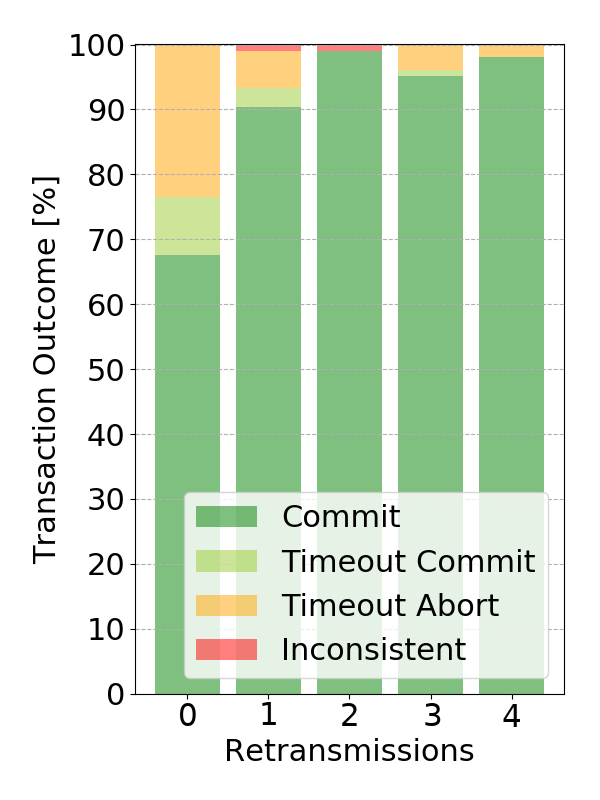}
  \caption{200ms slots}
  \label{fig:3pc-chaos-outcome-200ms}
\end{subfigure}
\caption{Agreement outcome of XPC 3PC-Chaos with varying slot duration in FlockLab.}
\label{fig:3pc-chaos-outcome}
\end{figure*}

\begin{figure*}[!ht]
% \vspace{-10pt}
\centering
\begin{subfigure}{.25\textwidth}
  \centering
  \includegraphics[scale=.20]{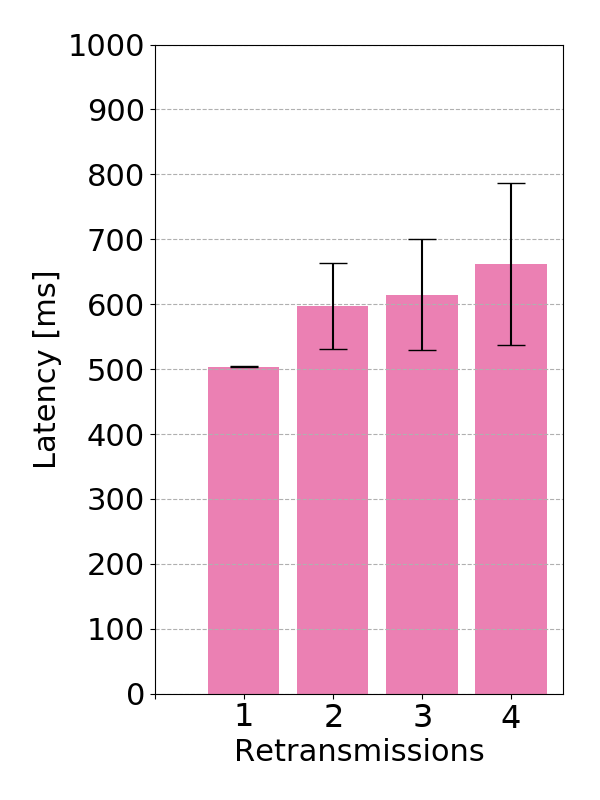}
  \caption{25ms slots}
  \label{fig:3pc-chaos-latency-25ms}
\end{subfigure}%
\begin{subfigure}{.25\textwidth}
  \centering
  \includegraphics[scale=.20]{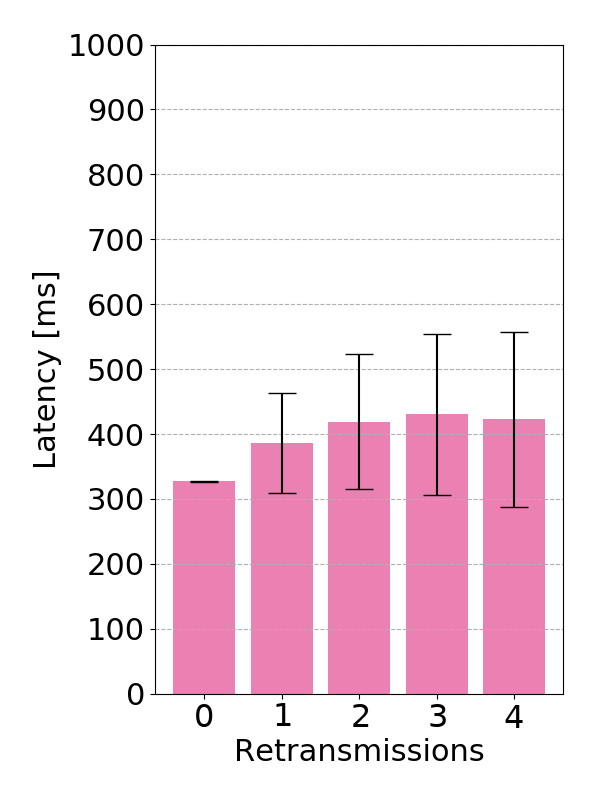}
  \caption{50ms slots}
  \label{fig:3pc-chaos-latency-50ms}
\end{subfigure}%
\begin{subfigure}{.25\textwidth}
  \centering
  \includegraphics[scale=.20]{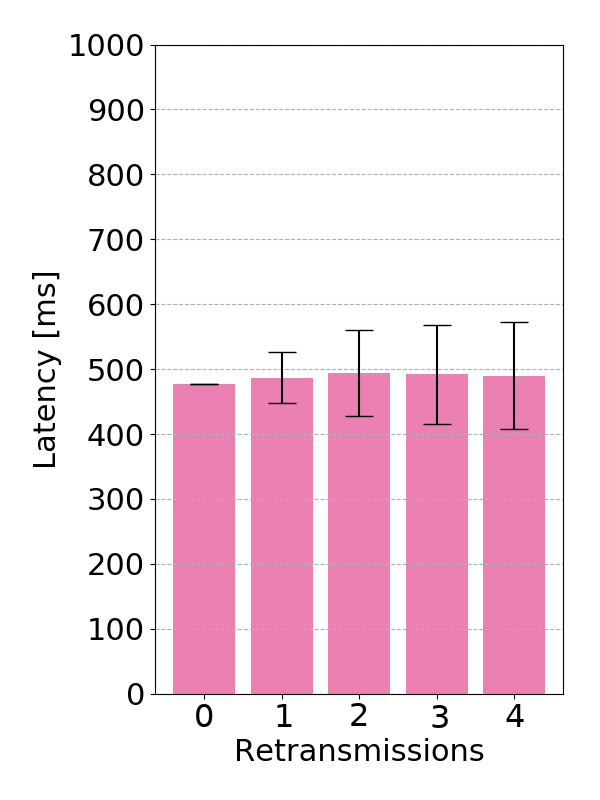}
  \caption{100ms slots}
  \label{fig:3pc-chaos-latency-100ms}
\end{subfigure}%
\begin{subfigure}{.25\textwidth}
  \centering
  \includegraphics[scale=.20]{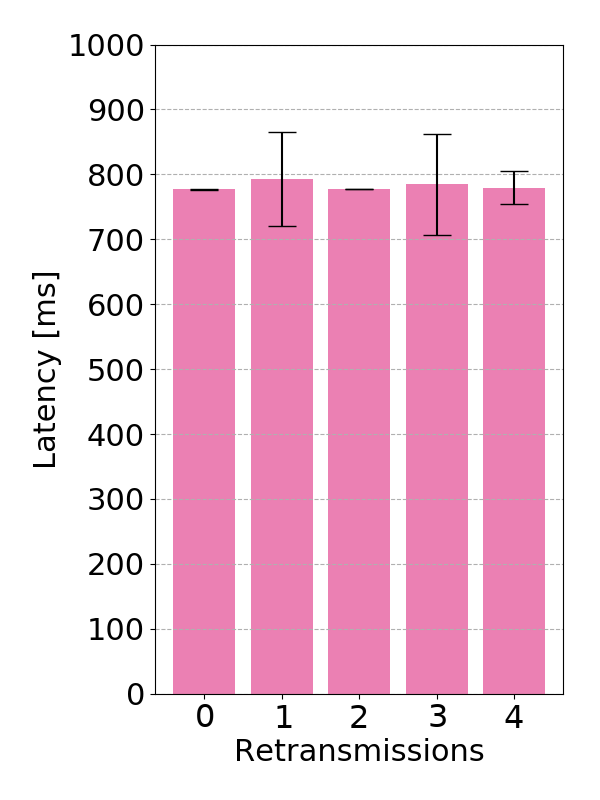}
  \caption{200ms slots}
  \label{fig:3pc-chaos-latency-200ms}
\end{subfigure}
\caption{Latency of XPC 3PC-Chaos with varying slot duration in FlockLab.}
\label{fig:3pc-chaos-latency}
\end{figure*}

The results show that 3PC-Chaos suffers from the same reliability and latency
concerns as 3PC-Glossy. The reliability of 3PC-Chaos (Figure
\ref{fig:3pc-chaos-outcome}) is worse than 2PC-Chaos. 
This is due to the addition of an extra Chaos round and the ``timeout commit''
that is a part of 3PC. Similar to 2PC-Chaos, above 90\% reliability 
can be achieved with 200ms transmission slots.

3PC-Chaos latency increases with the introduction of an additional
communication round, but remain lower than 3PC-Glossy (Figure
\ref{fig:3pc-chaos-latency}). Chaos is a fast ST primitive with a low overall
protocol execution time. Unfortunately, Chaos can be unreliable on low-power
multi-hop networks.

%%%%%%%%%%%%%%%%%%%%%%%%%%%%%%%%%%%%%%%%%%%%%%%%%%%%%%%%%%%%%%%%%%%%%%%%%%%%%%%%%%%%%%%%%%%%%%%%%%%%
%%%%%%%%%%%%%%%%%%%%%%%%%%%%%%%%%%%%%%%%%%%%%%%%%%%%%%%%%%%%%%%%%%%%%%%%%%%%%%%%%%%%%%%%%%%%%%%%%%%%
%%%%%%%%%%%%%%%%%%%%%%%%%%%%%%%%%%%%%%%%%%%%%%%%%%%%%%%%%%%%%%%%%%%%%%%%%%%%%%%%%%%%%%%%%%%%%%%%%%%%
\subsection{XPC using \textit{Hybrid}} \label{sec:xpc-hybrid}

In this section we evaluate XPC that harnesses both Chaos and Glossy into the same
protocol solving the
latency issues of Glossy and the reliability issues of Chaos. 

When utilising ST primitives it is very important to select the correct Chaos
slot duration and number of retransmissions. If the Chaos slot duration is too
short, few node replies will be received and many Glossy rounds will be
required.  If the initial Chaos slot duration is too long, time may be wasted
waiting for replies from nodes.

%%%%%%%%%%%%%%%%%%%%%%%%%%%%%%%%%%%%%%%%%%%%%%%%%%%%%%%%%%%%%%%%%%%%%%%%%%%%%%%%%%%%%%%%%%%%%%%%%%%%
\subsubsection{\textit{Hybrid} Two-Phase Commit}

Our results for 2PC-\textit{Hybrid} use the same Chaos slot duration used for
the evaluation of Chaos on its own. The slot duration only determines the
length of the first round of each protocol phase. The subsequent
retransmissions use Glossy.

\begin{figure*}[!ht]
\vspace{-10pt}
\centering
\begin{subfigure}{.25\textwidth}
  \centering
  \includegraphics[scale=.20]{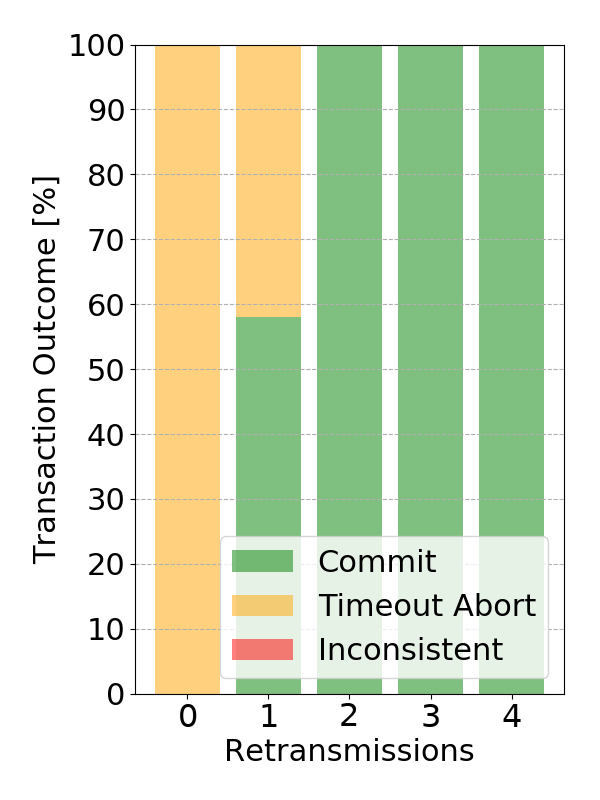}
  \caption{25ms slots}
  \label{fig:2pc-hybrid-outcome-25ms}
\end{subfigure}%
\begin{subfigure}{.25\textwidth}
  \centering
  \includegraphics[scale=.20]{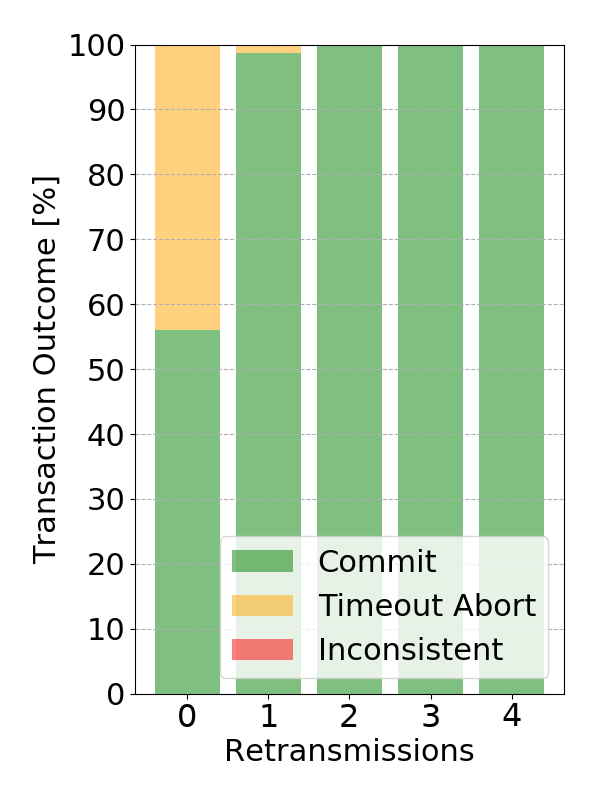}
  \caption{50ms slots}
  \label{fig:2pc-hybrid-outcome-50ms}
\end{subfigure}%
\begin{subfigure}{.25\textwidth}
  \centering
  \includegraphics[scale=.20]{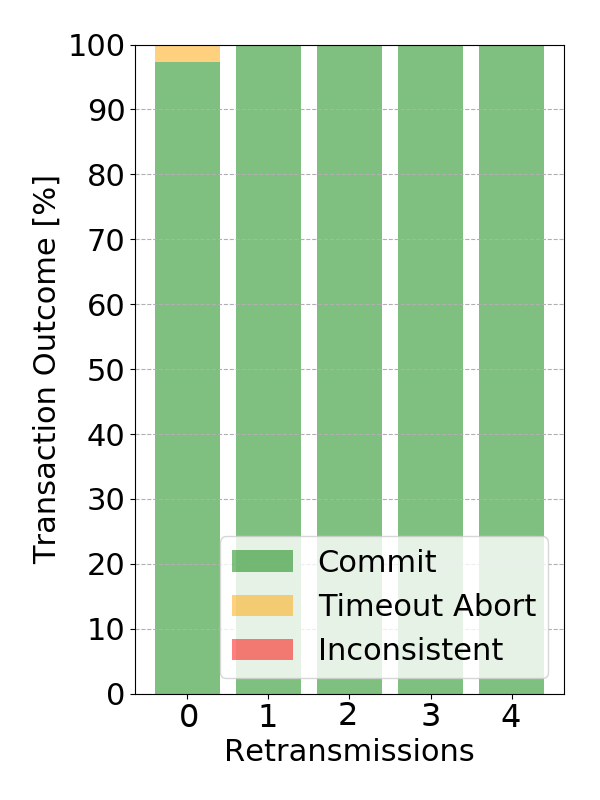}
  \caption{100ms slots}
  \label{fig:2pc-hybrid-outcome-100ms}
\end{subfigure}%
\begin{subfigure}{.25\textwidth}
  \centering
  \includegraphics[scale=.20]{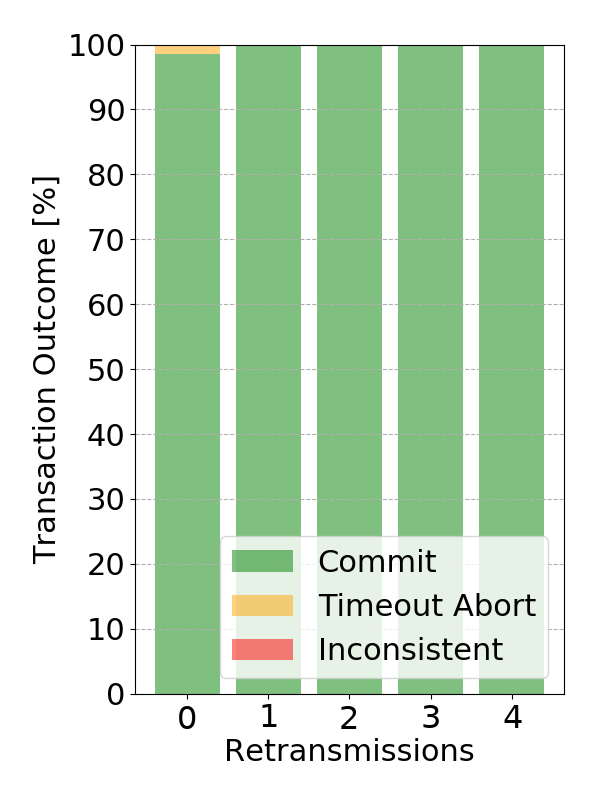}
  \caption{200ms slots}
  \label{fig:2pc-hybrid-outcome-200ms}
\end{subfigure}
\caption{Agreement outcome of XPC 2PC-\textit{Hybrid} with varying slot duration in FlockLab.}
\label{fig:2pc-hybrid-outcome}
\end{figure*}

\begin{figure*}[!ht]
\centering
\begin{subfigure}{.25\textwidth}
  \centering
  \includegraphics[scale=.20]{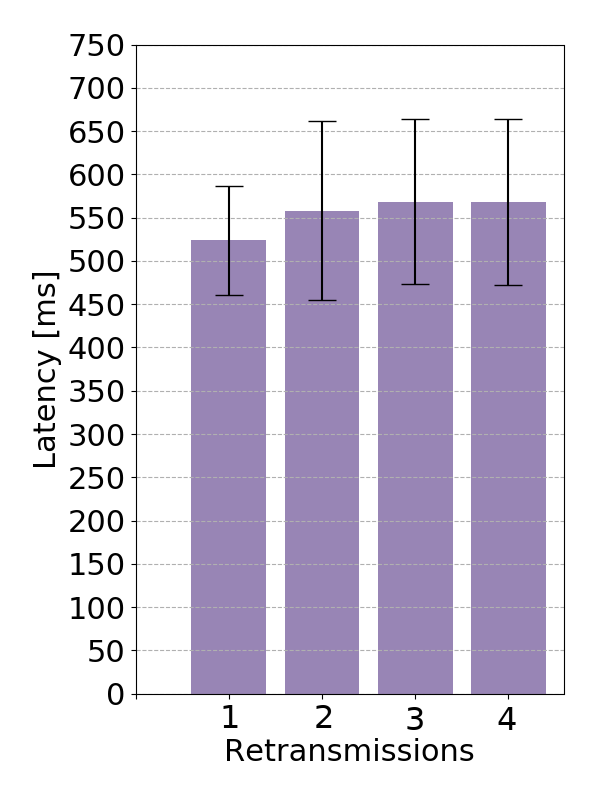}
  \caption{25ms slots}
  \label{fig:2pc-hybrid-latency-25ms}
\end{subfigure}%
\begin{subfigure}{.25\textwidth}
  \centering
  \includegraphics[scale=.20]{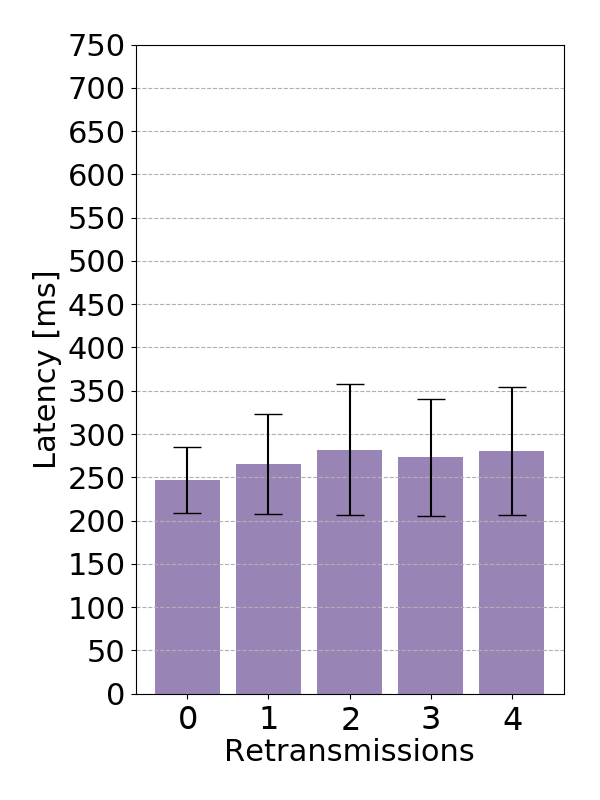}
  \caption{50ms slots}
  \label{fig:2pc-hybrid-latency-50ms}
\end{subfigure}%
\begin{subfigure}{.25\textwidth}
  \centering
  \includegraphics[scale=.20]{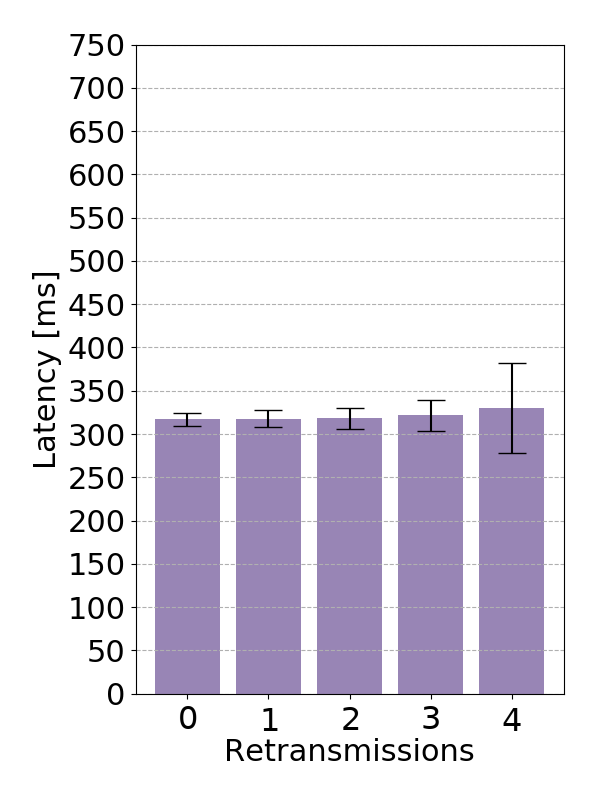}
  \caption{100ms slots}
  \label{fig:2pc-hybrid-latency-100ms}
\end{subfigure}%
\begin{subfigure}{.25\textwidth}
  \centering
  \includegraphics[scale=.20]{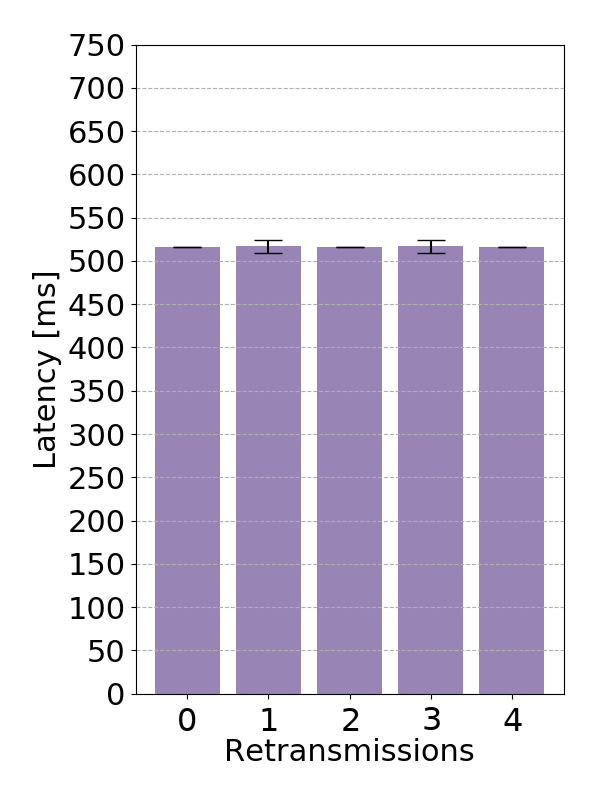}
  \caption{200ms slots}
  \label{fig:2pc-hybrid-latency-200ms}
\end{subfigure}
\caption{Latency of XPC 2PC-\textit{Hybrid} with varying slot duration in FlockLab.}
\label{fig:2pc-hybrid-latency}
\end{figure*}

We can see in Figure \ref{fig:2pc-hybrid-outcome} that the Glossy
retransmissions increase the protocol
reliability to 100\% for our experimental set-up. The latency (Figure \ref{fig:2pc-hybrid-latency}) is also
very close to that of 2PC-Chaos. 

%%%%%%%%%%%%%%%%%%%%%%%%%%%%%%%%%%%%%%%%%%%%%%%%%%%%%%%%%%%%%%%%%%%%%%%%%%%%%%%%%%%%%%%%%%%%%%%%%%%%
\subsubsection{\textit{Hybrid} Three-Phase Commit}

\begin{figure*}[!ht]
\vspace{-10pt}
\centering
\begin{subfigure}{.25\textwidth}
  \centering
  \includegraphics[scale=.20]{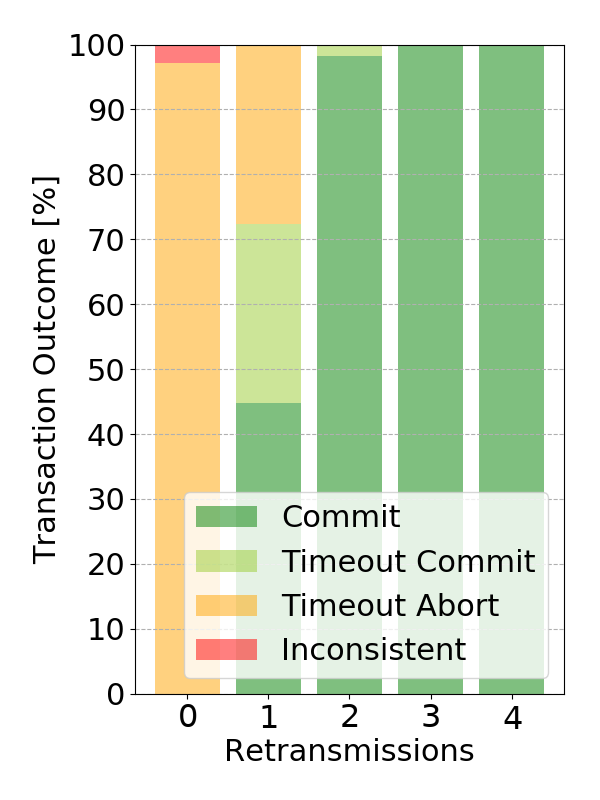}
  \caption{25ms slots}
  \label{fig:3pc-hybrid-outcome-25ms}
\end{subfigure}%
\begin{subfigure}{.25\textwidth}
  \centering
  \includegraphics[scale=.20]{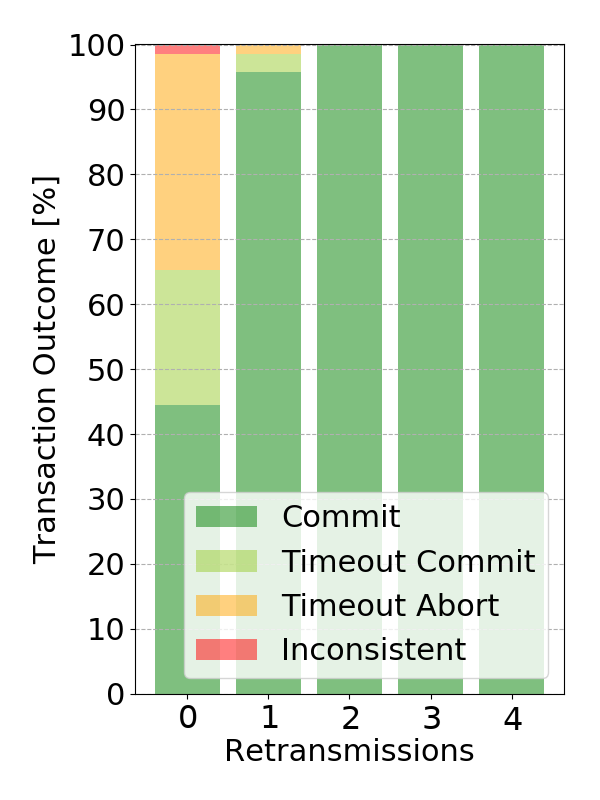}
  \caption{50ms slots}
  \label{fig:3pc-hybrid-outcome-50ms}
\end{subfigure}%
\begin{subfigure}{.25\textwidth}
  \centering
  \includegraphics[scale=.20]{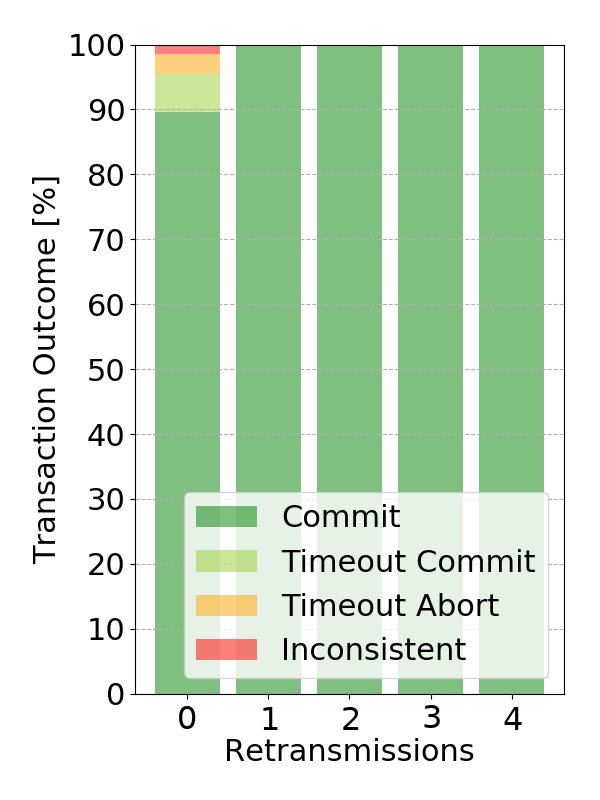}
  \caption{100ms slots}
  \label{fig:3pc-hybrid-outcome-100ms}
\end{subfigure}%
\begin{subfigure}{.25\textwidth}
  \centering
  \includegraphics[scale=.20]{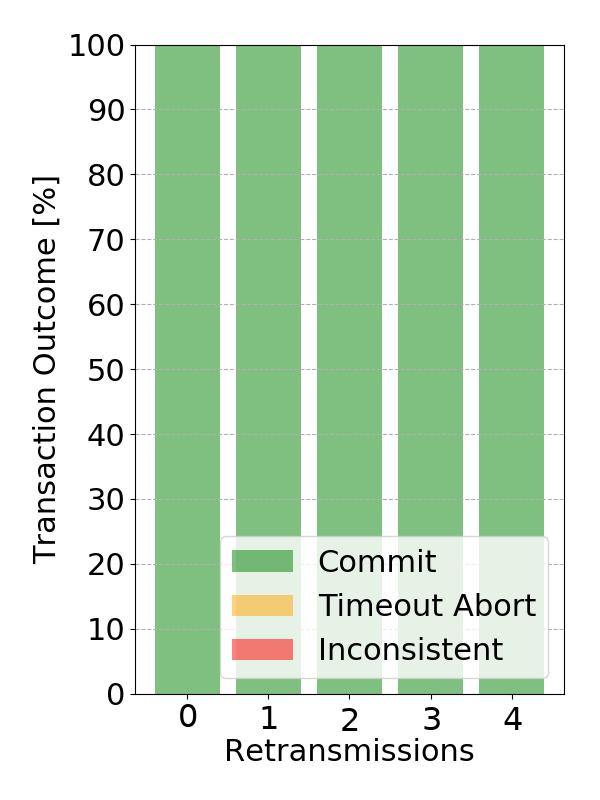}
  \caption{200ms slots}
  \label{fig:3pc-hybrid-outcome-200ms}
\end{subfigure}
\caption{Agreement outcome of XPC 3PC-\textit{Hybrid} with varying slot duration in FlockLab.}
\label{fig:3pc-hybrid-outcome}
\end{figure*}

\begin{figure*}[!ht]
\centering
\begin{subfigure}{.25\textwidth}
  \centering
  \includegraphics[scale=.20]{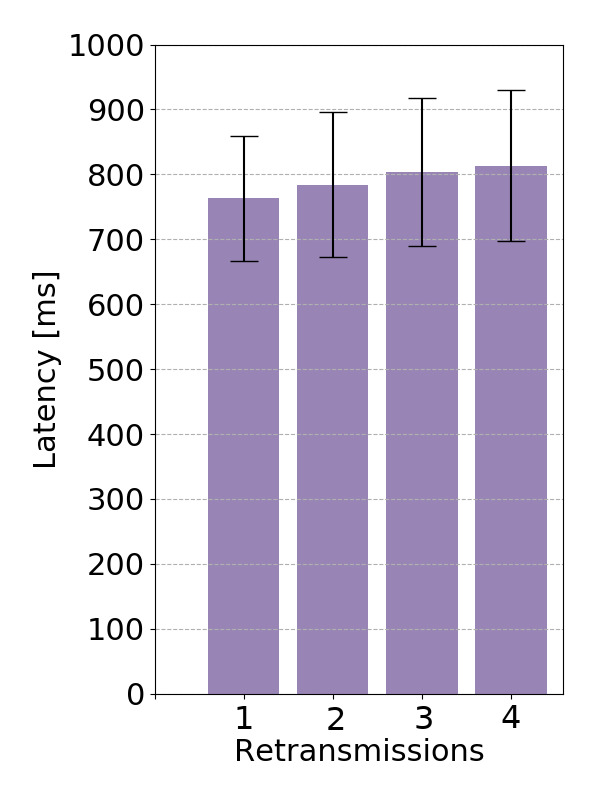}
  \caption{25ms slots}
  \label{fig:3pc-hybrid-latency-25ms}
\end{subfigure}%
\begin{subfigure}{.25\textwidth}
  \centering
  \includegraphics[scale=.20]{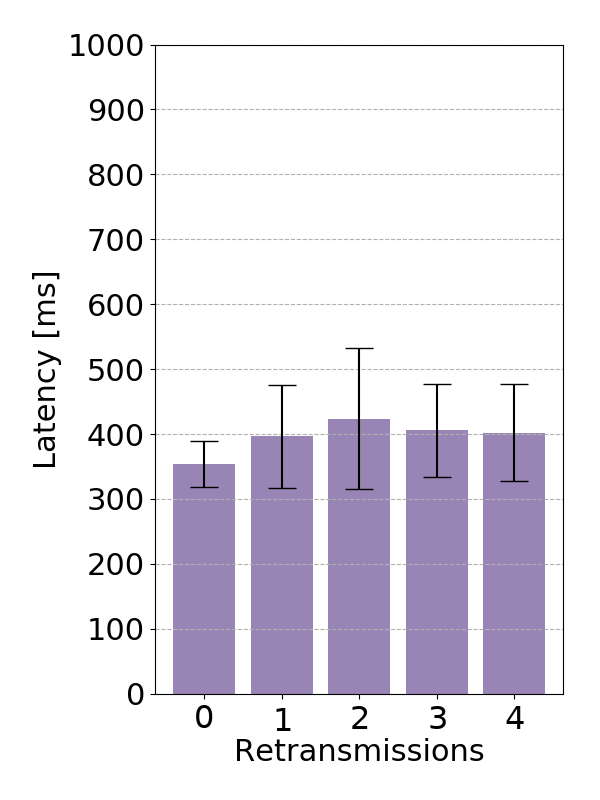}
  \caption{50ms slots}
  \label{fig:3pc-hybrid-latency-50ms}
\end{subfigure}%
\begin{subfigure}{.25\textwidth}
  \centering
  \includegraphics[scale=.20]{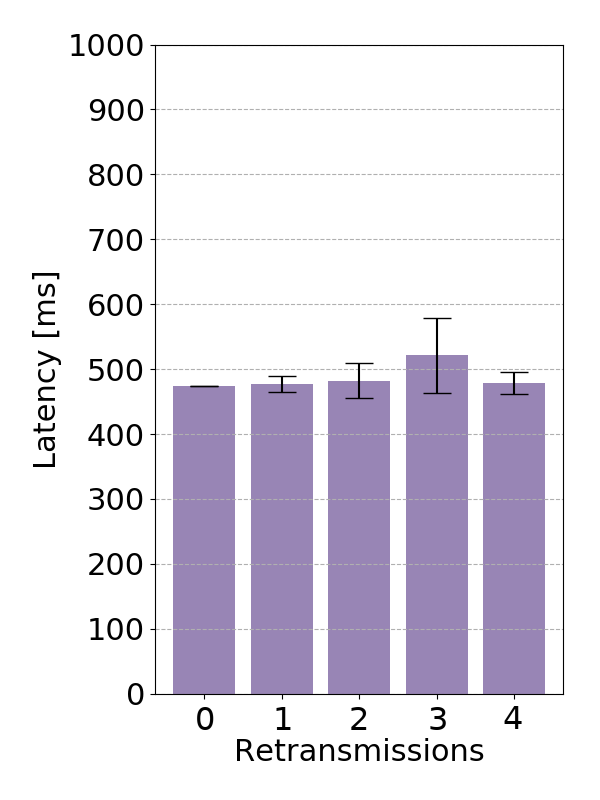}
  \caption{100ms slots}
  \label{fig:3pc-hybrid-latency-100ms}
\end{subfigure}%
\begin{subfigure}{.25\textwidth}
  \centering
  \includegraphics[scale=.20]{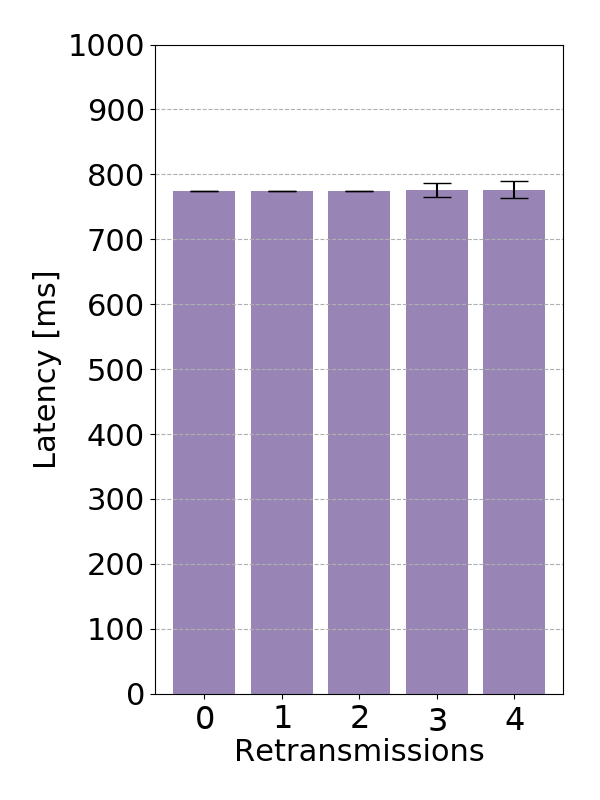}
  \caption{200ms slots}
  \label{fig:3pc-hybrid-latency-200ms}
\end{subfigure}
\caption{Latency of XPC 3PC-\textit{Hybrid} with varying slot duration in FlockLab.}
\label{fig:3pc-hybrid-latency}
\end{figure*}

The results for 3PC-\textit{Hybrid} (Figure \ref{fig:3pc-hybrid-outcome}) show that
agreement can be reached reliably with 4 or 5 retransmissions for any 
slot duration. A 100ms slot duration can reach 100\% reliability with 3
retransmissions. At 200ms high reliability can be achieved with no retransmissions.

The latency of 3PC-\textit{Hybrid} (see Figure \ref{fig:3pc-hybrid-latency}) is very
close to that of 3PC-Chaos. Our data shows that XPC using \textit{Hybrid} to
schedule ST primitives can be used to realise higher-level abstractions 
for use in synchronous WSNs with good performance and reliability.

%%%%%%%%%%%%%%%%%%%%%%%%%%%%%%%%%%%%%%%%%%%%%%%%%%%%%%%%%%%%%%%%%%%%%%%%%%%%%%%%%%%%%%%%%%%%%%%%%%%%
\subsection{Interference Analysis} \label{sec:eval-interference}

We focus our next set of experiments on the reliability of our \textit{Hybrid} 
%???HYBRID -- ARE WE GIVING ONE THING TWO NAMES <- should be clearer now???
under varying amounts of radio interference. Interference causes nodes to miss
broadcasted packets, and potentially desynchronise from the network (causing
transmission slots to be missed); protocol reliability must therefore be
analysed in the presence of network interference.

% as12015 C4.03
% > This whole paragraph
% We should now (or very very soon) have updated data which we can display. The data explores Glossy and Chaos as well so we can change the contents of this paragraph tor reflect that

Our previous experiments were done during times of low radio interference, i.e.
in good conditions. We established that \textit{Hybrid} out-performs both Glossy and
Chaos in such conditions.  In the next set of experiments we evaluate Chaos,
Glossy and \textit{Hybrid} with much greater radio interference. 

%The experiments were performed on the Flocklab testbed. Of the potential 27
%nodes available, 22 were used as the nodes of the network and the remaining 5
%nodes were used as jammers. The jamming nodes are selected as nodes that are
%physically close enough to another node to be able to jam its radio reception.
%The jamming nodes use JamLab \cite{jamlab}, a customizable off-the-shelf radio
%interference generation library for WSN motes. The following interference
%patterns (as discussed in \cite{jamlab}) were used to perform comparisons:

The experiments were performed on the Flocklab testbed. Of the 22 functioning
node, we choose between 1 and 8 nodes to inject interference into the network.
We increase the interference by using different interference models, and
different numbers of interfering nodes.
The interfering nodes were \{31, 20, 27, 28, 8, 6, 4, 3\}. These interfering
nodes were selected as nodes that are physically close enough to another node
to be able to jam its radio reception.  The jamming nodes use JamLab
\cite{jamlab}, a customizable off-the-shelf radio interference generation
library for WSN motes. JamLab provides a set of interference profiles 
(explained below), that are becoming an accepted way to evaluate WSN protocols
in comparable conditions. The following interference patterns (as discussed in
\cite{jamlab}) were used to perform comparisons:

%???THESE ARE BECOMING DEFACTO ACCEPTED TESTS FOR WSN INTERFERENCE  Explained???

\begin{enumerate}
    \item \textbf{Low Interference} is determined by background noise on the
    FlockLab testbed during night-time hours (9pm-6am). This models an ideal
    network deployment. 
    \item \textbf{High Interference} is determined by background noise on the
    FlockLab testbed during day-time hours (7am-8pm). It provides an estimate
    of average real-world conditions. 
    \item \textbf{WiFi Interference} is generated by JamLab and emulates the
    interference of non-saturated WiFi file transfers and radio streaming.
    \item \textbf{Microwave Interference} is generated by JamLab and emulates
    the periodic interference caused by microwave ovens over 802.15.4
    transmission channels.
\end{enumerate}

% as12015 C4.05
% > "each protocol phase is allowed up to 10 retransmissions before timing out and aborting"
% Technically 9 if you do not count the first round execution as a "retransmission"

All nodes in the network vote in favour of all proposed values (100\% agreement
rate) and each protocol phase is allowed up to 9 retransmissions before timing
out and aborting. The Chaos slot duration for Chaos and \textit{Hybrid} is $50$ms. The
interference experiments were run on Flocklab at a time when 22 node were
functional.

%\begin{table}[!ht]
%\begin{center}
%\scalebox{0.6}{%
%\begin{tabular}{?c|c?c|c?} \toprule
%    %% LOW INTERFERENCE %%
%    \multicolumn{2}{?c|}{\textbf{Parameters}} &
%    \multicolumn{2}{|c?}{\textbf{Protocol}} \\ \toprule
%    {\textbf{Name}} & {\textbf{Value}} & {\textbf{2PC}} & {\textbf{3PC}} \\ \midrule
%    Network Nodes & 22 & X & X \\ \midrule
%    Max Retransmmissions & 10 & X & X \\ \midrule
%    Chaos Round Length & 50ms & X & X \\ \midrule
%    Agreement Rate & 100\% & X & X \\ \midrule
%    %M-Slot Length & 35ms &  &  & X & X \\ \midrule
%    %$\Delta$ Q Size & 3 &  &  & X & X \\ \midrule
%    %Concurrent Initiators & 1 &  &  & & X \\ \bottomrule
%\end{tabular}
%}%scalebox
%\end{center}
%\caption{XPC configuration for Flocklab interference tests.} \label{table:eval-config}
%\end{table}

%\begin{table}[!ht]
%\begin{center}
%\scalebox{0.6}{%
%\begin{tabular}{?c|c?} \toprule
%    %% LOW INTERFERENCE %%
%    \multicolumn{2}{?c|}{\textbf{Parameters}} \\ \toprule 
%    %\multicolumn{2}{|c?}{\textbf{Protocol}} \\ \toprule
%    {\textbf{Name}} & {\textbf{Value}} \\ \midrule
%    Network Nodes & 22 \\ \midrule
%    Max Retransmmissions & 10 \\ \midrule
%    Chaos Round Length & 50ms \\ \midrule
%    Agreement Rate & 100\% \\ \midrule
%    %M-Slot Length & 35ms &  &  & X & X \\ \midrule
%    %$\Delta$ Q Size & 3 &  &  & X & X \\ \midrule
%    %Concurrent Initiators & 1 &  &  & & X \\ \bottomrule
%\end{tabular}
%}%scalebox
%\end{center}
%\caption{XPC configuration for Flocklab interference tests.} \label{table:eval-config}
%\end{table}

It is important to note that we consider WiFi and Microwave interference to both represent a 
high degree of radio interference. 

Both types of JamLab injected interference (WiFI and Microwave) were executed
during night-time hours to minimise other external interference. Our FlockLab
interference analysis uses the following:

% as12015 C4.06
% > "Chaos Round Coverage"
% Now "1st Round Coverage" or similar

\begin{itemize}
    \item \textbf{Interference Model}. Low, High, WiFi and Microwave
    interference models were tested and evaluated individually for each
    protocol.
    \item \textbf{Average Reliability}. Protocol reliability measures the rate
    at which all nodes in the network commit the proposed transaction
    consistently. \textit{If even just one node times out or aborts, the reliability is}
    \textit{scored as zero for the given round}.
    \item \textbf{Latency}. Latency measures the overall time from the
    beginning of an XPC transaction (i.e. when the Application layer is
    pre-empted) until the application is resumed at the end of the XPC round.
    Protocol latency is expected to increase with the interference.
    \item \textbf{First Round Coverage}. The First Round Coverage metric expresses
    the percentage of network nodes reached, on average, during the first
    dissemination of each phase, which are denoted as P1, P2 and P3 depending
    on the number of phases. 
    \item \textbf{Average Number of retransmissions}. This metric expresses the
    average number of retransmissions required for a protocol to switch to a
    subsequent stage. 
\end{itemize}

%%%%%%%%%%%%%%%%%%%%%%%%%%%%%%%%%%%%%%%%%%%%%%%%%%%%%%
\subsection{Single Jammer}

\input{Sections/flocklab-interference-extended.tex}

%%%%%%%%%%%%%%%%%%
%%%% END BIG TABLE Version 2!
%%%%%%%%%%%%%%%%%%

The results of 2PC-Glossy, 2PC-Chaos, 2PC-\textit{Hybrid} and 3PC-\textit{Hybrid}, evaluated under the four interference
models outlined above can be seen in Table \ref{table:evaluation-global}. The
Table presents the interference caused by one node with a variety of different
interference patterns.

An interesting result is the poor latency and reliability of 2PC-Chaos across
all experiments under these harsher (but realistic) conditions. This result is not surprising, and has been observed by others
using Chaos based protocols \cite{al2018competition}. A point needs to be made
about the difference in latency results for Chaos in these experiments when
compared to Figure \ref{fig:2pc-chaos-latency}. Here we measure the overall latency of
both committed and aborted (due to timeout) transactions.  In Figure
\ref{fig:2pc-chaos-latency} we only measure the latency of the commit
transactions. This shows us the potential benefits of combining Chaos and
Glossy in real world end-to-end execution a system in the presence of
interference. 

We make the following further observations about the data presented in Table
\ref{table:evaluation-global}.
\begin{itemize}
    \item \textbf{Reliability}. The protocols aside from Chaos across all interference models
    achieve a 100\% correct outcome. This data shows the high
    reliability and low latency of \textit{Hybrid}. Not only is the correct
    functionality of the protocols maintained in ideal (Low Interference) and
    normal (High Interference) network conditions, but it is also resilient 
    to network interference injected to cause packet loss and
    broadcast conflicts.
    \item \textbf{Latency}. As the interference models increase their
    disturbance over the channel, protocol latency's increase linearly.
    The stronger WiFi and Microwave radio interference causes longer latency
    for 3PC-\textit{Hybrid}. 
    \item \textbf{Chaos Coverage}. In spite of the problems with Chaos when
    used on its own, when combined with Glossy into \textit{Hybrid} it works very well.
    Being able to reach over 90\% of nodes during the first 50ms of each phase
    reduces latency; subsequently switching to reliable Glossy broadcasts is
    then able to compromise for Chaos' unpredictable termination time and lack
    of reliable detection of straggler nodes. 
    \item \textbf{Average retransmissions}. Similar to the analysis with
    latency and Chaos round coverage, the average number of phase
    retransmissions reflects the intensity of the channel's interference. As
    the interference increases all of the protocols require more
    retransmissions to maintain reliability.
\end{itemize}

%%%%%%%%%%%%%%%%%%%%%%%%%%%%%%%%%%%%%%%%%%%%%%%%%%%%%%
\subsection{Multiple Jammers}

% as12015 C4.07
% > "We only evaluate the 2PC-\textit{Hybrid} universal-voting protocol"
% WIll be incorrect if we add the additional data

We extend our analysis to consider the impact of increased interference in the network. The
purpose of this evaluation is to disrupt the network in degrees until we can
see complete failure. We do this using multiple nodes generating jamming
interference in the network. We select the microwave oven as an extreme form of
interference 
%???WHY PEOPLE TEND TO HAVE ONLY ONE MICROWAVE OVEN IN THEIR BUILDINGS - DOES THIS REPRESENT 
%OTHER SIMILAR THINGS AT 2.4 <-- Explained above???. 
By increasing the number nodes generating microwave oven
interference we create a more challenging communication environment for the
evaluation of the robustness of the protocol. It is important to note that all of the degrees 
of interference represented in this experiment are high. We consider that beyond three 
jamming nodes represents extreme interference that is beyond what would be expected of a normal
operational environment. Table
\ref{table:evaluation-multiference-extended} reports the results of executions with
between 2 and 8 interfering nodes.

\input{Sections/flocklab-multiference-extended.tex}
%%%%%%%%%%%%%%%%%%
%%%% END BIG TABLE 2 Version 2!
%%%%%%%%%%%%%%%%%%

%As shown in Table \ref{table:evaluation-multiference} high levels of
%interference will impact the protocol's reliability. With 4 interfering nodes
%2PC-\textit{Hybrid} only commits ~57\% of its transactions.  The
%remaining ~43\% time out due to missing replies. A decrease in reliability also
%affects latency and average retransmissions. The average retransmissions with 4
%interfering nodes is more than 6. The traces show that with failures,
%the protocol reaches the maximum number of retransmissions for this test.

The results in Table \ref{table:evaluation-multiference-extended} show that
2PC-\textit{Hybrid} is faster than 2PC-Glossy for small amounts of interference. At one
to two interfering nodes 2PC-Glossy has higher
latency than 2PC-\textit{Hybrid} because it does not have the initial Chaos flood used
by \textit{Hybrid} to efficiently flood the network with data. We can see that the first
transmission coverage is higher and the average retransmissions are lower for
Glossy, but the latency is higher.

At three interfering
nodes the reliability of 2PC-Glossy and 2PC-\textit{Hybrid} begin to degrade. We still
see very similar reliability for both. At this point the initial Chaos round
reaches fewer nodes and both protocols rely on Glossy floods. With four
interfering nodes 2PC-Glossy is more reliable than 2PC-\textit{Hybrid}. This occurs
because both are now reliant upon Glossy, and 2PC-Glossy has 10 glossy
retransmissions while 2PC-\textit{Hybrid} has 9 glossy retransmissions. This trend
continues as the number of interfering nodes increases. With six interfering
nodes the protocols has essentially failed. None have a reliability above 10\%. 

The unreliability of Chaos rounds are clearly seen with the performance of
2PC-Chaos. At four interfering nodes 2PC-Chaos has completely failed. We also
see that Chaos has a very high count of average retransmissions. 

3PC-\textit{Hybrid} behaves similarly to 2PC-\textit{Hybrid}, with a 50\% extra latency due to the extra
phase. From the experiment traces we can see that the 1st dissemination phase
usually has the highest retransmissions. This is probably caused by 
nodes finding it harder to capture the control packet for the next round while under
interference. At 8 jamming nodes we note that 3PC-\textit{Hybrid} completely fails to receive any packets 
for the first Chaos transmit of the second and third phases. We also see the total number of 
retransmissions at their maximum value.

%%%%%%%%%%%%%%%%%%%%%%%%%%%%%%%%%%%%%%%%%%%%%%%%%%%%%%%%%%%%%%%%%%%%%%%%%%%%%%%%%%%%%%%%%%%%%%%%%%%%
%\section{Evaluation}

%%%%%%%%%%%%%%%%%%%%%%%%%%%%%%%%%%%%%%%%%%%%%%%%%%%%%%%%%%%%%%%%%%%%%%%%%%%%%%%%%%%%%%%%%%%%%%%%%%%%

%In this section we have validated the reliability, latency and broadcast
%efficiency claims made in this paper about the \textit{Hybrid} ST approach enabled by
%our programming abstraction XPC. Our implementation of two-phase commit and
%three-phase commit are robust and resilient to high levels of network
%interference.

%% file: Sections/flocklab-interference-extended.tex
\begin{table}[!ht]
\begin{center}
%\resizebox{\textwidth}{!}{%
\scalebox{0.55}{
\begin{tabular}{|c|cccc|} \toprule
    %% LOW INTERFERENCE %%
    {\textbf{Low Interference}} & {\textbf{Reliability (\%)}} & {\textbf{Latency (ms)}} & {\textbf{Chaos Coverage (\%)}}& {\textbf{Avg. Retr.}}  \\ \midrule
     \multirow{2}{*}{2PC Glossy} & \multirow{2}{*}{100.00} & \multirow{2}{*}{564.43} & \textbf{P1}: 99.72 [G] & \textbf{P1}: 1.08 \\
           & & & \textbf{P2}: 99.81 [G] & \textbf{P2}: 1.03 \\ \midrule
     \multirow{2}{*}{2PC Chaos} & \multirow{2}{*}{95.77} & \multirow{2}{*}{287.00} & \textbf{P1}: 96.37 [C] & \textbf{P1}: 1.27 \\
           & & & \textbf{P2}: 96.36 [C] & \textbf{P2}: 1.41 \\ \midrule
     \multirow{2}{*}{2PC \textit{Hybrid}} & \multirow{2}{*}{100.00} & \multirow{2}{*}{244.17} & \textbf{P1}: 97.98 [C] & \textbf{P1}: 1.13 \\
            & & & \textbf{P2}: 97.14 [C] & \textbf{P2}: 1.21 \\ \midrule
     \multirow{3}{*}{3PC \textit{Hybrid}} & \multirow{3}{*}{100.00} & \multirow{3}{*}{409.79} & \textbf{P1}: 96.29 [C] & \textbf{P1}: 1.41 \\
            & & & \textbf{P2}: 97.57 [C] & \textbf{P2}: 1.35 \\
            & & & \textbf{P3}: 97.28 [C] & \textbf{P3}: 1.37 \\ \midrule
    %% HIGH INTERFERENCE %%
    {\textbf{High Interference}} & {\textbf{Reliability (\%)}} & {\textbf{Latency (ms)}} & {\textbf{Chaos Coverage (\%)}}& {\textbf{Avg. Retr.}}  \\ \midrule
     \multirow{2}{*}{2PC Glossy} & \multirow{2}{*}{100.00} & \multirow{2}{*}{580.95} & \textbf{P1}: 98.95 [G] & \textbf{P1}: 1.15 \\
           & & & \textbf{P2}: 98.54 [G] & \textbf{P2}: 1.25 \\ \midrule
     \multirow{2}{*}{2PC Chaos} & \multirow{2}{*}{89.09} & \multirow{2}{*}{481.66} & \textbf{P1}: 94.77 [C] & \textbf{P1}: 2.71 \\
           & & & \textbf{P2}: 96.13 [C] & \textbf{P2}: 2.46 \\ \midrule
     \multirow{2}{*}{2PC \textit{Hybrid}} & \multirow{2}{*}{100.00} & \multirow{2}{*}{284.79} & \textbf{P1}: 96.64 [C] & \textbf{P1}: 1.63 \\
            & & & \textbf{P2}: 97.26 [C] & \textbf{P2}: 1.46 \\ \midrule
     \multirow{3}{*}{3PC \textit{Hybrid}} & \multirow{3}{*}{100.00} & \multirow{3}{*}{465.52} & \textbf{P1}: 94.19 [C] & \textbf{P1}: 1.63 \\
            & & & \textbf{P2}: 95.57 [C] & \textbf{P2}: 1.70 \\
            & & & \textbf{P3}: 95.88 [C] & \textbf{P3}: 1.68 \\ \midrule
    %% WIFI INTERFERENCE %%
    {\textbf{Wifi Interference}} & {\textbf{Reliability (\%)}} & {\textbf{Latency (ms)}} & {\textbf{Chaos Coverage (\%)}}& {\textbf{Avg. Retr.}}  \\ \midrule
     \multirow{2}{*}{2PC Glossy} & \multirow{2}{*}{100.00} & \multirow{2}{*}{628.63} & \textbf{P1}: 98.04 [G] & \textbf{P1}: 1.60 \\
           & & & \textbf{P2}: 98.04 [G] & \textbf{P2}: 1.58 \\ \midrule
     \multirow{2}{*}{2PC Chaos} & \multirow{2}{*}{56.76} & \multirow{2}{*}{875.74} & \textbf{P1}: 89.53 [C] & \textbf{P1}: 3.94 \\
           & & & \textbf{P2}: 87.95 [C] & \textbf{P2}: 5.51 \\ \midrule
     \multirow{2}{*}{2PC \textit{Hybrid}} & \multirow{2}{*}{100.00} & \multirow{2}{*}{357.29} & \textbf{P1}: 91.36 [C] & \textbf{P1}: 1.78 \\
            & & & \textbf{P2}: 91.18 [C] & \textbf{P2}: 1.72 \\ \midrule
     \multirow{3}{*}{3PC \textit{Hybrid}} & \multirow{3}{*}{100.00} & \multirow{3}{*}{499.00} & \textbf{P1}: 93.72 [C] & \textbf{P1}: 1.71 \\
            & & & \textbf{P2}: 91.44 [C] & \textbf{P2}: 1.62 \\
            & & & \textbf{P3}: 93.89 [C] & \textbf{P3}: 1.67 \\ \midrule
    %% MICROWAVE INTERFERENCE %%
    {\textbf{Microwave}} & {\textbf{Reliability (\%)}} & {\textbf{Latency (ms)}} & {\textbf{Chaos Coverage (\%)}}& {\textbf{Avg. Retr.}}  \\ \midrule
     \multirow{2}{*}{2PC Glossy} & \multirow{2}{*}{100.00} & \multirow{2}{*}{593.33} & \textbf{P1}: 98.89 [G] & \textbf{P1}: 1.29 \\
           & & & \textbf{P2}: 98.96 [G] & \textbf{P2}: 1.29 \\ \midrule
     \multirow{2}{*}{2PC Chaos} & \multirow{2}{*}{70.97} & \multirow{2}{*}{889.92} & \textbf{P1}: 86.36 [C] & \textbf{P1}: 5.51 \\
           & & & \textbf{P2}: 89.18 [C] & \textbf{P2}: 4.51 \\ \midrule
     \multirow{2}{*}{2PC \textit{Hybrid}} & \multirow{2}{*}{100.00} & \multirow{2}{*}{355.11} & \textbf{P1}: 92.99 [C] & \textbf{P1}: 1.73 \\
            & & & \textbf{P2}: 92.44 [C] & \textbf{P2}: 1.86 \\ \midrule
     \multirow{3}{*}{3PC \textit{Hybrid}} & \multirow{3}{*}{100.00} & \multirow{3}{*}{507.28} & \textbf{P1}: 93.13 [C] & \textbf{P1}: 1.69 \\
            & & & \textbf{P2}: 91.83 [C] & \textbf{P2}: 1.76 \\
            & & & \textbf{P3}: 93.07 [C] & \textbf{P3}: 1.64 \\ \midrule
\end{tabular}}
\end{center}

\caption{Comparisons of XPC protocols for different interference patterns (1
jamming node).} \label{table:evaluation-global}
\end{table}

%% file: Sections/flocklab-multiference-extended.tex
\begin{table}[!b]
\vspace{-10pt}
\begin{center}
\scalebox{0.55}{%
%\resizebox{\textwidth}{!}{%
\begin{tabular}{|c|cccc|} \toprule
    %% MICROWAVE 1 NODES %%
    {\textbf{1 Jamming Node}} & {\textbf{Reliability (\%)}} & {\textbf{Latency (ms)}} & {\textbf{1\textsuperscript{st} Tx Coverage (\%)}}& {\textbf{Avg. Retr.}}  \\ \midrule
    \multirow{2}{*}{2PC Glossy} & \multirow{2}{*}{100.00} & \multirow{2}{*}{593.33} & \textbf{P1}: 98.89 [G] & \textbf{P1}: 1.29 \\
          & & & \textbf{P2}: 98.96 [G] & \textbf{P2}: 1.29 \\
          \midrule
    \multirow{2}{*}{2PC Chaos} & \multirow{2}{*}{70.97} & \multirow{2}{*}{889.92} & \textbf{P1}: 86.36 [C] & \textbf{P1}: 5.51 \\
          & & & \textbf{P2}: 89.18 [C] & \textbf{P2}: 4.51 \\
          \midrule
    \multirow{2}{*}{2PC \textit{Hybrid}} & \multirow{2}{*}{100.00} & \multirow{2}{*}{355.11} & \textbf{P1}: 92.99 [C] & \textbf{P1}: 1.73 \\
          & & & \textbf{P2}: 92.44 [C] & \textbf{P2}: 1.86 \\
          \midrule
    \multirow{2}{*}{3PC \textit{Hybrid}} & \multirow{2}{*}{100.00} & \multirow{2}{*}{507.28} & \textbf{P1}: 93.13 [C] & \textbf{P1}: 1.69 \\
          & & & \textbf{P2}: 91.83 [C] & \textbf{P2}: 1.76 \\
          & & & \textbf{P3}: 93.07 [C] & \textbf{P3}: 1.64 \\
          \midrule
    %% MICROWAVE 2 NODES %%
    {\textbf{2 Jamming Nodes}} & {\textbf{Reliability (\%)}} & {\textbf{Latency (ms)}} & {\textbf{1\textsuperscript{st} Tx Coverage (\%)}}& {\textbf{Avg. Retr.}}  \\ \midrule
    \multirow{2}{*}{2PC Glossy} & \multirow{2}{*}{100.00} & \multirow{2}{*}{787.45} & \textbf{P1}: 91.28 [G] & \textbf{P1}: 2.56 \\
          & & & \textbf{P2}: 91.28 [G] & \textbf{P2}: 2.85 \\
          \midrule
    \multirow{2}{*}{2PC Chaos} & \multirow{2}{*}{49.59} & \multirow{2}{*}{1030.10} & \textbf{P1}: 60.64 [C] & \textbf{P1}: 6.30 \\
          & & & \textbf{P2}: 80.71 [C] & \textbf{P2}: 5.89 \\
          \midrule
    \multirow{2}{*}{2PC \textit{Hybrid}} & \multirow{2}{*}{100.00} & \multirow{2}{*}{648.85} & \textbf{P1}: 82.67 [C] & \textbf{P1}: 3.05 \\
          & & & \textbf{P2}: 82.87 [C] & \textbf{P2}: 3.05 \\
          \midrule
    \multirow{2}{*}{3PC \textit{Hybrid}} & \multirow{2}{*}{100.00} & \multirow{2}{*}{976.25} & \textbf{P1}: 77.43 [C] & \textbf{P1}: 3.37 \\
          & & & \textbf{P2}: 81.66 [C] & \textbf{P2}: 3.09 \\
          & & & \textbf{P3}: 81.29 [C] & \textbf{P3}: 2.96 \\
          \midrule
    %% MICROWAVE 3 NODES %%
    {\textbf{3 Jamming Nodes}} & {\textbf{Reliability (\%)}} & {\textbf{Latency (ms)}} & {\textbf{1\textsuperscript{st} Tx Coverage (\%)}}& {\textbf{Avg. Retr.}}  \\ \midrule
    \multirow{2}{*}{2PC Glossy} & \multirow{2}{*}{94.14} & \multirow{2}{*}{969.45} & \textbf{P1}: 85.62 [G] & \textbf{P1}: 3.77 \\
          & & & \textbf{P2}: 86.50 [G] & \textbf{P2}: 3.60 \\
          \midrule
    \multirow{2}{*}{2PC Chaos} & \multirow{2}{*}{34.31} & \multirow{2}{*}{987.03} & \textbf{P1}: 73.75 [C] & \textbf{P1}: 6.45 \\
          & & & \textbf{P2}: 78.16 [C] & \textbf{P2}: 5.39 \\
          \midrule
    \multirow{2}{*}{2PC \textit{Hybrid}} & \multirow{2}{*}{95.00} & \multirow{2}{*}{841.74} & \textbf{P1}: 78.28 [C] & \textbf{P1}: 4.07 \\
          & & & \textbf{P2}: 79.62 [C] & \textbf{P2}: 3.87 \\
          \midrule
    \multirow{2}{*}{3PC \textit{Hybrid}} & \multirow{2}{*}{97.79} & \multirow{2}{*}{1270.12} & \textbf{P1}: 78.95 [C] & \textbf{P1}: 3.95 \\
          & & & \textbf{P2}: 77.60 [C] & \textbf{P2}: 4.05 \\
          & & & \textbf{P3}: 77.79 [C] & \textbf{P3}: 3.85 \\
          \midrule
    %% MICROWAVE 4 NODES %%
    {\textbf{4 Jamming Nodes}} & {\textbf{Reliability (\%)}} & {\textbf{Latency (ms)}} & {\textbf{1\textsuperscript{st} Tx Coverage (\%)}}& {\textbf{Avg. Retr.}}  \\ \midrule
    \multirow{2}{*}{2PC Glossy} & \multirow{2}{*}{52.10} & \multirow{2}{*}{1123.97} & \textbf{P1}: 84.66 [G] & \textbf{P1}: 6.37 \\
          & & & \textbf{P2}: 86.39 [G] & \textbf{P2}: 5.68 \\
          \midrule
    \multirow{2}{*}{2PC Chaos} & \multirow{2}{*}{0.20} & \multirow{2}{*}{1066.59} & \textbf{P1}: 56.12 [C] & \textbf{P1}: 8.96 \\
          & & & \textbf{P2}: 67.55 [C] & \textbf{P2}: 7.94 \\
          \midrule
    \multirow{2}{*}{2PC \textit{Hybrid}} & \multirow{2}{*}{46.94} & \multirow{2}{*}{1134.01} & \textbf{P1}: 76.71 [C] & \textbf{P1}: 7.27 \\
          & & & \textbf{P2}: 78.82 [C] & \textbf{P2}: 6.34 \\
          \midrule
    \multirow{2}{*}{3PC \textit{Hybrid}} & \multirow{2}{*}{40.63} & \multirow{2}{*}{1480.84} & \textbf{P1}: 74.22 [C] & \textbf{P1}: 7.23 \\
          & & & \textbf{P2}: 78.95 [C] & \textbf{P2}: 5.58 \\
          & & & \textbf{P3}: 75.60 [C] & \textbf{P3}: 5.80 \\
          \midrule
    %% MICROWAVE 5 NODES %%
    {\textbf{5 Jamming Nodes}} & {\textbf{Reliability (\%)}} & {\textbf{Latency (ms)}} & {\textbf{1\textsuperscript{st} Tx Coverage (\%)}}& {\textbf{Avg. Retr.}}  \\ \midrule
    \multirow{2}{*}{2PC Glossy} & \multirow{2}{*}{23.01} & \multirow{2}{*}{1274.33} & \textbf{P1}: 77.07 [G] & \textbf{P1}: 8.13 \\
          & & & \textbf{P2}: 83.22 [G] & \textbf{P2}: 7.33 \\
          \midrule
    \multirow{2}{*}{2PC Chaos} & \multirow{2}{*}{0.00} & \multirow{2}{*}{1061.64} & \textbf{P1}: 56.47 [C] & \textbf{P1}: 9.13 \\
          & & & \textbf{P2}: 68.99 [C] & \textbf{P2}: 7.10 \\
          \midrule
    \multirow{2}{*}{2PC \textit{Hybrid}} & \multirow{2}{*}{14.94} & \multirow{2}{*}{1291.59} & \textbf{P1}: 68.35 [C] & \textbf{P1}: 8.87 \\
          & & & \textbf{P2}: 75.47 [C] & \textbf{P2}: 7.41 \\
          \midrule
    \multirow{2}{*}{3PC \textit{Hybrid}} & \multirow{2}{*}{13.79} & \multirow{2}{*}{1506.73} & \textbf{P1}: 63.49 [C] & \textbf{P1}: 8.74 \\
          & & & \textbf{P2}: 79.80 [C] & \textbf{P2}: 7.52 \\
          & & & \textbf{P3}: 77.68 [C] & \textbf{P3}: 6.67 \\
          \midrule
    %% MICROWAVE 6 NODES %%
    {\textbf{6 Jamming Nodes}} & {\textbf{Reliability (\%)}} & {\textbf{Latency (ms)}} & {\textbf{1\textsuperscript{st} Tx Coverage (\%)}}& {\textbf{Avg. Retr.}}  \\ \midrule
    \multirow{2}{*}{2PC Glossy} & \multirow{2}{*}{2.66} & \multirow{2}{*}{1192.78} & \textbf{P1}: 72.24 [G] & \textbf{P1}: 8.87 \\
          & & & \textbf{P2}: 83.31 [G] & \textbf{P2}: 7.61 \\
          \midrule
    \multirow{2}{*}{2PC Chaos} & \multirow{2}{*}{0.00} & \multirow{2}{*}{1041.92} & \textbf{P1}: 53.16 [C] & \textbf{P1}: 9.27 \\
          & & & \textbf{P2}: 63.65 [C] & \textbf{P2}: 7.67 \\
          \midrule
    \multirow{2}{*}{2PC \textit{Hybrid}} & \multirow{2}{*}{1.56} & \multirow{2}{*}{1169.43} & \textbf{P1}: 68.55 [C] & \textbf{P1}: 8.78 \\
          & & & \textbf{P2}: 78.62 [C] & \textbf{P2}: 7.11 \\
          \midrule
    \multirow{2}{*}{3PC \textit{Hybrid}} & \multirow{2}{*}{1.94} & \multirow{2}{*}{1277.95} & \textbf{P1}: 67.54 [C] & \textbf{P1}: 9.14 \\
          & & & \textbf{P2}: 76.36 [C] & \textbf{P2}: 7.64 \\
          & & & \textbf{P3}: 78.38 [C] & \textbf{P3}: 7.14 \\
          \midrule
    %% MICROWAVE 7 NODES %%
    {\textbf{7 Jamming Nodes}} & {\textbf{Reliability (\%)}} & {\textbf{Latency (ms)}} & {\textbf{1\textsuperscript{st} Tx Coverage (\%)}}& {\textbf{Avg. Retr.}}  \\ \midrule
    \multirow{2}{*}{2PC Glossy} & \multirow{2}{*}{1.15} & \multirow{2}{*}{1231.18} & \textbf{P1}: 66.15 [G] & \textbf{P1}: 9.62 \\
          & & & \textbf{P2}: 76.99 [G] & \textbf{P2}: 8.35 \\
          \midrule
    \multirow{2}{*}{2PC Chaos} & \multirow{2}{*}{0.00} & \multirow{2}{*}{1054.03} & \textbf{P1}: 46.10 [C] & \textbf{P1}: 9.80 \\
          & & & \textbf{P2}: 60.00 [C] & \textbf{P2}: 10.00 \\
          \midrule
    \multirow{2}{*}{2PC \textit{Hybrid}} & \multirow{2}{*}{0.39} & \multirow{2}{*}{1230.55} & \textbf{P1}: 61.19 [C] & \textbf{P1}: 9.59 \\
          & & & \textbf{P2}: 74.45 [C] & \textbf{P2}: 8.30 \\
          \midrule
    \multirow{2}{*}{3PC \textit{Hybrid}} & \multirow{2}{*}{0.80} & \multirow{2}{*}{1293.27} & \textbf{P1}: 59.80 [C] & \textbf{P1}: 9.66 \\
          & & & \textbf{P2}: 73.67 [C] & \textbf{P2}: 8.82 \\
          & & & \textbf{P3}: 76.30 [C] & \textbf{P3}: 8.52 \\
          \midrule
    %% MICROWAVE 8 NODES %%
    {\textbf{8 Jamming Nodes}} & {\textbf{Reliability (\%)}} & {\textbf{Latency (ms)}} & {\textbf{1\textsuperscript{st} Tx Coverage (\%)}}& {\textbf{Avg. Retr.}}  \\ \midrule
    \multirow{2}{*}{2PC Glossy} & \multirow{2}{*}{0.00} & \multirow{2}{*}{1252.55} & \textbf{P1}: 58.79 [G] & \textbf{P1}: 9.74 \\
          & & & \textbf{P2}: 76.40 [G] & \textbf{P2}: 7.00 \\
          \midrule
    \multirow{2}{*}{2PC Chaos} & \multirow{2}{*}{0.00} & \multirow{2}{*}{1048.42} & \textbf{P1}: 41.44 [C] & \textbf{P1}: 9.82 \\
          & & & \textbf{P2}: 37.33 [C] & \textbf{P2}: nan \\
          \midrule
    \multirow{2}{*}{2PC \textit{Hybrid}} & \multirow{2}{*}{0.00} & \multirow{2}{*}{1336.88} & \textbf{P1}: 51.40 [C] & \textbf{P1}: 9.75 \\
          & & & \textbf{P2}: 68.26 [C] & \textbf{P2}: 8.00 \\
          \midrule
    \multirow{2}{*}{3PC \textit{Hybrid}} & \multirow{2}{*}{0.00} & \multirow{2}{*}{1300.07} & \textbf{P1}: 50.49 [C] & \textbf{P1}: 10.00 \\
          & & & \textbf{P2}: 0.00 [C] & \textbf{P2}: 10.00 \\
          & & & \textbf{P3}: 0.00 [C] & \textbf{P3}: 10.00 \\
          \midrule
\end{tabular}}
\end{center}

\caption{Comparisons of XPC protocols with multiple jamming nodes (Microwave
Interference).}\label{table:evaluation-multiference-extended}
\end{table}

%% file: Sections/Discussion_Limitations_FurtherWork.tex
\section{Limitations and Further Work}

%This paper looks at the approaches to robustness and reliability that can be
%obtained by treating Synchronous Transmissions as communication primitives. It
%has looked at how, with the right abstraction, these communication primitives
%can enable protocols that were previously difficult to implement on WSN with
%unreliable communication links. 

%???SAY SOMETHING ABOUT SCALE
%???HERE  -  DO  WE  WORRY  ABOUT  THAT  NOT  BEING
%???LARGE - SAY WHY WE DON’T WORk -- Done
The limitations of this work are typical of those in this field. It is
difficult to control the radio environment of a remote WSN testbed. We hold
that our best-effort experiments with the injection of radio interference do
tell us something useful about the resilience of the hybrid use of Glossy and
Chaos. A more controlled environment could have given us more precise results.

\textit{Hybrid} also suffers from the same issues of scale shared by Glossy and Chaos. 
Glossy needs an individual network-wide flood for each node. Chaos uses a 
control frame that contains one bit for each node in the network. Both of these limit
the size of network that each can be used on.

We are developing this work as part of a large smart city wide deployment of sensors in (name 
hidden for double blind review) to support the management of lifts, waste pipelines and pollution 
monitoring applications. This will enable both parameter updates and code updates as a result.  
For further work, we would also like to extend the use of XPC and the \textit{Hybrid} ST approach to see 
what further communication protocols could be supported, or what
new ones could be developed. We would also like to explore a way to incorporate
the use of multiple channels to increase resilience.

%% file: Sections/Conclusion.tex
\section{Conclusion}

In this paper we present X Phase Commit(XPC) for the implementation of atomic
commit protocols using Synchronous Transmissions, and \textit{Hybrid}. We describe the design of
XPC and reference implementations of two-phase commit and
three-phase commit using two of the most common ST primitives, Glossy and
Chaos. We also present \textit{Hybrid}, a way to use both Glossy and Chaos
to provide fast and reliable
flooding. We used XPC and our reference implementations to assess the
latency and reliability of Glossy, Chaos, and \textit{Hybrid} for both the two-phase
commit and three-phase commit transactional protocols.  

% as12015 C6.01 > "The XPC library is available online at GITHUB" Technically
% currently only on GitLab and it's private, should we make it public? How
% should we separate the WISP/WIMP?

Our testbed evaluation showed that \textit{Hybrid} enabled by XPC has a lower latency
than Glossy on its own, and is more reliable than Chaos on its own. We evaluate
evaluated Glossy, Chaos, and \textit{Hybrid} with increasing levels of network radio
interference and saw that under low to moderate interference \textit{Hybrid} was as
reliable as Glossy but with a lower latency. The XPC library is available
online at GITLAB.
%https://gitlab.doc.ic.ac.uk/xpc.